\documentstyle[preprint,aps,epsf,floats,12pt]{revtex}

\begin{document}

\vskip 1.5 cm

\preprint{\tighten \vbox{\hbox{} }}

\title{Angular Distribution of Charming $B \to V V $ Decays and
Time Evolution Effects}
\author{Cheng-Wei Chiang\footnote{E-mail:
chengwei@andrew.cmu.edu}}

\address{
Department of Physics,
Carnegie Mellon University,
Pittsburgh, Pennsylvania 15213}

\maketitle

{\tighten
\begin{abstract}
Angular distributions of a $B$ meson decaying into two vector mesons
are discussed with emphasis on time evolution effects on the complete
set of amplitude bilinears.  Time integrated quantities are suggested
to observe substantial CP violation in decays with charm quarks in the
final state particles.  Relations among the nine observables at $t=0$
are found to be useful for a consistency check of experimentally
extracted quantities.  Numerical estimates of the nine observables are
made using form factor models and the assumption of the factorization
hypothesis.  Branching ratio asymmetries for $B_u^+ \to D^{*+} \bar
D^{*0}$ and $B_d \to D^{*+} D^{*-}$ can be as large as $-3\%$ and
$-4\%$, respectively.
\end{abstract}
}

%%%%%%%%%%%%%%%%%%%%%%%%%%%%%%%%%%%%%%%%%%%%%%%%%%%%%%%%%%%%%%%

\newpage

\section{Introduction}

The decays of the $B$ meson into two vector mesons $B \to
V_1\,+\,V_2$, either with charm quarks in the final state particles,
such as $B \to J/\Psi\,\rho$, or with particles without charm quarks,
such as $B \to \rho\,K^*$, have been calculated in many models
\cite{DDLR96,DDF99,KP92PRD,KP92PLB,KPS94,KMP92,F99,LSS99,CT99}.  The
time evolution effects in neutral $B$ meson decays are also discussed
in \cite{DDF99,F99}.  In this work, we would like to extend the
general discussion on time evolving observables and to emphasize the
charming decays in numerical analysis.

One major advantage of analyzing $B \to VV$ decays is that the
interference of CP-even and CP-odd final states appear in the angular
distributions.  These interference terms provide good opportunities to
observe CP or T violating effects.  Since it is possible to measure
all nine observables in certain decays \cite{CW99}, the physically
interesting quantities such as $\beta$ and $\eta$ can be determined
from experiments given sufficient statistics.  In addition, relations
among the nine observables provide a consistency check for the
amplitude bilinears obtained experimentally.

The decay amplitude involves the hadronic matrix element of a $B$
meson decaying to two vector mesons through a weak current, which at
present can not be calculated from first principles.  Thus, our
numerical evaluation of these observables at $t=0$ is based upon the
assumptions of factorization and no final state interactions and form
factor models, where the updated Wilson coefficients \cite{CY99} are
used.

This paper is organized as follows: In Section II, we review the
observables in the angular distributions of $B \to VV$ decays.  In
Section III, we derive the time-dependent formulas for the observables
and list the complete results in the Appendix.  Section IV considers
the situation of no time evolution or at $t=0$.  The case with no
strong phases is discussed in Section V.  Results of single weak
amplitude decays are presented in Section VI, wherein CP asymmetries
are also extensively discussed.  In Section VII, we present the
numerical estimation of the nine observables.  We summarize this paper
in Section VIII.

%%%%%%%%%%%%%%%%%%%%%%%%%%%%%%%%%%%%%%%%%%%%%%%%%%%%%%%%%%%%%%%

\section{Observables and Angular Distributions in $B \to
VV$ Decays}

To extract the $CP$-odd and $CP$-even or $T$-odd and $T$-even
components more easily, the angular distribution is often written in
the transversity basis.  Let us define the amplitude of $B \to V_1
V_2$ in the rest frame of $V_1$.  According to their polarization
combinations, the amplitude can be decomposed into \cite{DDLR96}
\begin{equation}
\label{ampl}
A(B \to V_1 V_2) = 
   A_0\,{\epsilon}^{*L}_{V_1} {\epsilon}^{*L}_{V_2}
 - \frac{A_{\|}}{\sqrt{2}}\,
     \vec {\epsilon}^{*T}_{V_1} \cdot \vec {\epsilon}^{*T}_{V_2}
 - i \frac{A_{\perp}}{\sqrt{2}}\,
     \vec {\epsilon}^*_{V_1} \times \vec {\epsilon}^*_{V_2}
     \cdot \hat{\bf p}, 
\end{equation}
and similarly for $\bar B \to \bar V_1 \bar V_2$.  In
Eq.~(\ref{ampl}), $\vec {\epsilon}_{V_1}$ and $\vec {\epsilon}_{V_2}$
are the unit polarization vectors of $V_1$ and $V_2$, respectively.
$\hat{\bf p}$ is the unit vector along the direction of motion of
$V_2$ in the rest frame of $V_1$, $\epsilon_{V_i}^{*L} \equiv \vec
{\epsilon}^*_{V_i} \cdot \hat{\bf p}$ and $\vec {\epsilon}^{*T}_{V_i} =
\vec {\epsilon}^*_{V_i} - \epsilon_{V_i}^{*L} \hat{\bf p}$.  It is
easy to see that $A_{\perp}$ is odd under the parity transformation
because of the appearance of $\vec {\epsilon}^*_{V_1} \times \vec
{\epsilon}^*_{V_2} \cdot \hat{\bf p}$, whereas $A_0$ and $A_{\|}$ are
even.

The nine observables in the squared amplitude $A^*A$ are \cite{CW99}
\begin{eqnarray}
\label{bilinears}
K_1(t) = |A_0(t)|^2, & \qquad
K_4(t) = Re \left[ A_0^*(t)A_{\|}(t) \right], & \qquad
L_4(t) = Im \left[ A_0^*(t)A_{\|}(t) \right], \nonumber \\
K_2(t) = |A_{\|}(t)|^2, & \qquad
K_5(t) = Im \left[ A_0^*(t)A_{\perp}(t) \right], & \qquad
L_5(t) = Re \left[ A_0^*(t)A_{\perp}(t) \right], \\
K_3(t) = |A_{\perp}(t)|^2, & \qquad
K_6(t) = Im \left[ A_{\|}^*(t)A_{\perp}(t) \right], & \qquad
L_6(t) = Re \left[ A_{\|}^*(t)A_{\perp}(t) \right]. \nonumber
\end{eqnarray}
So we have
\begin{eqnarray}
A^*(t)A(t) &=& 
   K_1(t)X_1(\Omega)\,+\,K_2(t)X_2(\Omega)\,
+\,K_3(t)X_3(\Omega)\, \nonumber \\
&& \qquad
+\,K_4(t)X_4(\Omega)\,
+\,L_5(t)X_5(\Omega)\,+\,L_6(t)X_6(\Omega)\, \nonumber \\
&& \qquad
+\,L_4(t)Y_4(\Omega)\,+\,K_5(t)Y_5(\Omega)\,
+\,K_6(t)Y_6(\Omega), \nonumber
\end{eqnarray}
where the quantities $X_i(\Omega)$ and $Y_i(\Omega)$ represent
polarizations or polarization correlations of the final vector mesons
and $\Omega$ stands for the angles of the outgoing particles.

In general, the angular distribution of the decay in the transversity
basis can be written as
\begin{equation}
\label{angdist}
\frac{d^3\Gamma(t)}{d\cos\theta_1 d\cos\theta_2 d\phi} =
\sum_{i} K_i(t) f_i(\theta_1,\theta_2,\phi),
\end{equation}
where $K_i$'s are the amplitude bilinears that contain the dynamics
and generally evolve with time, and $f_i(\theta_1,\theta_2,\phi)$ are the
corresponding angular distribution functions.

One can classify the decays into three types of processes according to
the properties of the final product particles as follows:

{\it Type I}\,: For the case in which the decays of $V_1$ and $V_2$ are
both into two pseudoscalar mesons, one can immediately translate the
tensor correlations into angular distributions \cite{CW99}.  The
normalized angular distribution of the decays $B \to V_1 (\to P_1
P_1^{\prime}) \, V_2 (\to P_2 P_2^{\prime})$, where $P_1^{({\prime})}$
and $P_2^{({\prime})}$ denote pseudoscalar mesons, is:
\begin{eqnarray}
\label{BtoVPPVPP}
\frac{1}{\Gamma_0} \frac{d^3\Gamma(t)}{d\cos\theta_1 d\cos\theta_2 d\phi}
& = & 
\frac{9}{8 \pi} \biggl \{
  \frac{K_1(t)}{\Gamma_0} \cos^2\theta_1 \cos^2\theta_2
+ \frac{K_2(t)}{2 \Gamma_0} 
       \sin^2\theta_1 \sin^2\theta_2 \cos^2\phi 
\biggr. \nonumber \\
&& 
+ \frac{K_3(t)}{2 \Gamma_0} 
       \sin^2\theta_1 \sin^2\theta_2 \sin^2\phi
+ \frac{K_4(t)}{2 \sqrt{2} \Gamma_0}
       \sin 2\theta_1 \sin 2\theta_2 \cos\phi \\
&&
\biggl.
- \frac{K_5(t)}{2 \sqrt{2} \Gamma_0}
       \sin 2\theta_1 \sin 2\theta_2 \sin \phi
- \frac{K_6(t)}{2 \Gamma_0}
       \sin^2\theta_1 \sin^2\theta_2 \sin 2\phi
\biggr \}. \nonumber
\end{eqnarray}
Here $\theta_1$ ($\theta_2$) is the angle between the $P_1$ ($P_2$)
three-momentum vector in the $V_1 ( V_2)$ rest frame and the $V_1$
($V_2$) three-momentum vector defined in the $B$ rest frame, and
$\phi$ is the angle between the normals to the planes defined by $P_1
P_1^\prime$ and $P_2 P_2^\prime$, in the $B$ rest frame.  Examples of
such decays are $B^+ \to \bar D^{*} (\to \bar D^0 \pi^0) \rho^+ (\to
\pi^+ \pi^0)$, $B_d \to D^{*-} (\to \bar D^0 \pi^-) \rho^+ (\to
\pi^+ \pi^0)$, and $B_d \to D^{*-} (\to \bar D^0 \pi^-) D^{*+} (\to
D^0 \pi^+)$.

{\it Type II}\,: For the case of the decay $B \to V_1\,(\to
P_1\,P_1^{\prime}) \, V_2\,(\to l^+\,l^-)$, suppose we observe that
$l^-$ is a right-handed particle and comes out in the direction
$\vec{k_2}=(\sin\theta_2\cos\phi,\sin\theta_2\sin\phi,\cos\theta_2)$
and the momentum of $P_1$, $\vec{k_1}=(\sin\theta_1,0,\cos\theta_1)$
with angles defined in the same fashion as in the previous type, we
have instead the differential angular distribution \cite{CW99}:
\begin{eqnarray}
\label{BtoVPPVLL}
\frac1{\Gamma_0}\frac{d^3\Gamma}{d\cos\theta_1 d\cos\theta_2 d\phi}
&=&
\frac{9}{16\pi\Gamma_0}
\biggl \{
  K_1 \cos^2\theta_1 \sin^2\theta_2
+ \frac{K_2}{2}
     \left(
       \sin^2\theta_1 \cos^2\theta_2 \cos^2\phi 
       + \sin^2\theta_1 \sin^2\phi
     \right)
\biggr. \nonumber \\
&& 
+ \frac{K_3}{2}
     \left(
       \sin^2\theta_1 \cos^2\theta_2 \sin^2\phi 
       + \sin^2\theta_1 \cos^2\phi
     \right)
+ \frac{K_4}{2 \sqrt{2}}
       \sin 2\theta_1 \sin 2\theta_2 \cos\phi \nonumber \\
&&
- \frac{K_5}{2 \sqrt{2}}
       \sin 2\theta_1 \sin 2\theta_2 \sin \phi
- \frac{K_6}{2}
       \sin^2\theta_1 \sin^2\theta_2 \sin 2\phi \nonumber \\
&&
\biggl.
+ \frac{L_4}{\sqrt{2}}
       \sin 2\theta_1 \sin \theta_2 \sin\phi
- \frac{L_5}{\sqrt{2}}
       \sin 2\theta_1 \sin \theta_2 \cos\phi
+ \frac{L_6}{2}
     \sin^2\theta_1 \cos\theta_2
\biggr \}
\end{eqnarray}
To obtain the result for the other possible final state with a
left-handed outgoing $l^-$, one only needs to flip the signs of $L_4$,
$L_5$, and $L_6$.  The muon polarization is equal to the sum of the
terms $L_4$, $L_5$, $L_6$ divided by the sum of the other 6 terms.
For the case of $L_6$ it is seen that the polarization does not vanish
after integrating over $\theta_1$ and $\phi$ and so the observation
can be made without observing the $V_1$ decay.  Such decay modes are
$B_u^+ \to J/\Psi (\to l^+ l^-) K^{*+} (\to \pi^0 K^+)$, $B_u^+ \to
J/\Psi (\to l^+ l^-) \rho^+ (\to \pi^+ \pi^0)$, $B_d \to J/\Psi (\to
l^+ l^-) K^* (\to \pi K)$, $B_d \to J/\Psi (\to l^+ l^-) \rho^0 (\to
\pi \pi)$, $B_d \to J/\Psi (\to l^+ l^-) \omega (\to \pi^+ \pi^-
\pi^0)$, $B_s \to J/\Psi (\to l^+ l^-) \bar K^* (\to \pi \bar K)$, and
$B_d \to J/\Psi (\to l^+ l^-) \phi (\to K^+ K^-)$.  Although $\omega$
decays into three pions, they are still correlated so that one can
pick the normal direction to the decay plane formed by the three pions
in the $\omega$ rest frame to define the direction $\theta_2$ and
$\phi$.

Although the $B \to V (\to PP) V (\to P \gamma)$ modes have a
different decay pattern from that of $B \to V (\to PP) V (\to l^+
l^-)$, they share the same differential angular distribution, with
the direction of $l^-$ in the latter case replaced by that of
$\gamma$.  For instance, for the decay with a right-handed circularly
polarized photon in the final state, the angular distribution is the
same as Eq.~(\ref{BtoVPPVLL}).  Such examples are $B_u^+ \to D_s^{*+}
(\to D_s^+ \gamma) \bar D^{*0} (\to \bar D^0 \pi)$, $B_d \to D_s^{*+}
(\to D_s^+ \gamma) D^{*-} (\to \bar D^0 \pi^-)$, $B_s \to D_s^{*-}
(\to D_s^- \gamma) D^{*+} (\to D^0 \pi^+)$.  If one does not measure
the polarization of the product particles, the angular distribution
would be the one by doubling Eq.~(\ref{BtoVPPVLL}) and eliminating the
$L_{4,5,6}$ terms.

{\it Type III}\,: Next we consider the decay $B \to V (\to P \gamma) V
(\to P \gamma)$.  Since it is experimentally impractical to measure
the polarizations of both photons in the final state, we just give
here the differential angular distribution with no polarization
measured:
\begin{eqnarray}
\label{BtoVPrVPr}
\frac1{\Gamma_0}\frac{d^3\Gamma}{d\cos\theta_1 d\cos\theta_2 d\phi}
&=&
\frac{9}{8\pi\Gamma_0}
\biggl \{
  K_1 \sin^2\theta_1 \sin^2\theta_2 \biggr. \nonumber \\
&&
+ \frac{K_2}{2}
     \left(
       \cos^2\theta_1 \cos^2\theta_2 \cos^2\phi 
       + \cos^2\theta_1 \sin^2\phi
       + \cos^2\theta_2 \sin^2\phi
       + \cos^2\phi
     \right) \nonumber \\
&& 
+ \frac{K_3}{2}
     \left(
       \cos^2\theta_1 \cos^2\theta_2 \sin^2\phi 
       + \cos^2\theta_1 \cos^2\phi
       + \cos^2\theta_2 \cos^2\phi
       + \sin^2\phi
     \right) \nonumber \\
&&
- \frac{K_4}{2 \sqrt{2}}
       \sin 2\theta_1 \sin 2\theta_2 \cos\phi
+ \frac{K_5}{2 \sqrt{2}}
       \sin 2\theta_1 \sin 2\theta_2 \sin \phi \nonumber \\
&&
\biggl.
+ \frac{K_6}{2}
       \sin^2\theta_1 \sin^2\theta_2 \sin 2\phi
\biggr \}
\end{eqnarray}
%

%%%%%%%%%%%%%%%%%%%%%%%%%%%%%%%%%%%%%%%%%%%%%%%%%%%%%%%%%%%%%%%

\section{Time Evolution of the Amplitude Bilinears}

The time evolution of an arbitrary neutral $B$ meson state $a |B^0 (t)
\rangle +b |\bar B^0(t) \rangle$ is governed by the Schr\"{o}dinger equation
\begin{equation}
i \frac{d}{dt} 
   \left( \begin{array}{c}
      a(t) \\ b(t)
   \end{array} \right)
= {\mathcal{H}}
   \left( \begin{array}{c}
      a(t) \\ b(t)
   \end{array} \right).
\end{equation}
If we write the mass eigenstates, $|B_{L,H} \rangle$, with eigenvalues
$m_{L,H}-\frac{i}{2} \Gamma_{L,H}$ in terms of $|B^0 \rangle$ and
$|\bar B^0 \rangle$ as
\begin{equation}
|B_{L,H} \rangle = p \, |B^0 \rangle \pm q \, |\bar B^0 \rangle,
\end{equation}
then the time evolutions of $B^0$ and $\bar B^0$ are
\begin{eqnarray}
\label{neutralBEvolve}
| B^0(t) \rangle &=&
   g_+(t) | B^0(0) \rangle
 + \frac{q}{p} \, g_-(t) | \bar B^0(t) \rangle, \nonumber \\
| \bar B^0(t) \rangle &=&
   \frac{p}{q} \, g_-(t) | B^0(0) \rangle
 + g_+(t) | \bar B^0(t) \rangle,
\end{eqnarray}
where
\begin{equation}
\label{evolvefcn}
g_{\pm}(t) = 
   \frac{1}{2} \left( e^{-i m_L t} e^{-\frac1{2} \Gamma_L t} \pm
   e^{-i m_H t} e^{-\frac1{2} \Gamma_H t} \right).
\end{equation}

Suppose $|f_{\eta} \rangle$ is a state with definite CP property,
namely, $CP |f_{\eta} \rangle = \eta_i |f_{\eta} \rangle$ for
$i=1,2,3$ and $\eta=0,\|,\perp$, respectively.  The CP eigenvalues
$\eta_1=\eta_2=+1$ and $\eta_3=-1$.  Suppose we write the decay matrix
element of $B^0$ decaying into the final states $f_{\eta}$ at time
$t=0$ as
\begin{equation}
\label{decaymatrix}
A_{\eta}(0) \equiv \langle f_{\eta}|B^0(0) \rangle
   = Y_{CKM}^T \, e^{i \theta_{\eta}}(T_{\eta}+P_{\eta} \, e^{i \phi_w}
   e^{i \delta_{\eta}}).
\end{equation}
$Y_{CKM}^T$ is the overall CKM factors appearing in the amplitudes.
$\theta_{\eta}$ are the factored strong phases of $A_{\eta}$, but only
the relative phases are essential.  Conventionally, we take
$\theta_{\bot}=0$.  $T_{\eta}$ and $P_{\eta}$ are the absolute values
of two types of amplitudes that differ by a relative weak phase
$\phi_w$ and a relative strong phase $\delta_{\eta}$.  We will refer
to them by ``tree'' and ``penguin'' amplitudes, respectively.
Similarly, for the CP conjugate mode we have
\begin{equation}
\label{conjugatedecaymatrix}
\bar A_{\eta}(0) \equiv \langle \bar f_{\eta}|\bar B^0(0) \rangle
   = {Y_{CKM}^T}^* \, e^{i \theta_{\eta}}(T_{\eta}+P_{\eta} \, e^{-i \phi_w}
   e^{i \delta_{\eta}}).
\end{equation}
Here we may assume that $|f_{\eta} \rangle$ and $|\bar f_{\eta}
\rangle$ are the same state that both $|B^0 \rangle$ and $|\bar B^0
\rangle$ can decay into ({\it e.g.}, $B^0, \bar B^0 \to
J/\Psi\,\phi,\,D^{*+}\,D^{*-}$).  They can also be conjugate states so
that only $|B^0 \rangle$ (or $|B^+ \rangle$) can decay into $|f_{\eta}
\rangle$ and only $|\bar B^0 \rangle$ (or $|B^- \rangle$) to $|\bar
f_{\eta} \rangle$.  According to the time evolution, the decay
amplitude at time $t$ would be
\begin{equation}
\label{evolve1}
A_{\eta}(t) = \langle f_{\eta}|B^0(t) \rangle
= A_{\eta}(0) \left[ g_+(t) + \eta_i \, \lambda_{\eta} g_-(t) \right],
\end{equation}
where
\begin{equation}
\label{lambda}
\lambda_{\eta} = \frac{q}{p} \frac{{Y_{CKM}^T}^*}{Y_{CKM}^T}
   \frac{T_{\eta}+P_{\eta} \, e^{-i \phi_w} e^{i \delta_{\eta}}}
        {T_{\eta}+P_{\eta} \, e^{i \phi_w} e^{i \delta_{\eta}}}.
\end{equation}
It is convenient to define a phase $\phi$ by
\begin{equation}
\label{phi}
e^{i \phi} \equiv \frac{q}{p} \frac{{Y_{CKM}^T}^*}{Y_{CKM}^T},
\end{equation}
and
\begin{equation}
\label{randi}
R_{\eta} \equiv Re 
  \left[
    \frac{T_{\eta}+P_{\eta} \, e^{-i \phi_w} e^{i \delta_{\eta}}}
         {T_{\eta}+P_{\eta} \, e^{i \phi_w} e^{i \delta_{\eta}}}
  \right], \, \,
I_{\eta} \equiv Im 
  \left[
    \frac{T_{\eta}+P_{\eta} \, e^{-i \phi_w} e^{i \delta_{\eta}}}
         {T_{\eta}+P_{\eta} \, e^{i \phi_w} e^{i \delta_{\eta}}}
  \right].
\end{equation}
Note that $R_{\eta}^2 + I_{\eta}^2 = 1$ if and only if
$\delta_{\eta}\, , \phi_w=0$ (mod $\pi$).  If either (i) no nontrivial
relative weak phase ($0$ or $\pi$), (ii) negligible tree contributions
($T_{\eta} \simeq 0$), or (iii) negligible penguin contributions
($P_{\eta} \simeq 0$) happens, then $R_{\eta} = 1$ and $I_{\eta} = 0$,
apart from a possible overall phase.

With the above definitions, one can get, for example, the time
evolving $|A_{\eta}(t)|^2$ as follows:
\begin{eqnarray}
\label{ampsquare}
&& |A_{\eta}(t)|^2 = |A_{\eta}(0)|^2 e^{-\Gamma t}
  \left \{ 
    \frac{1+R_{\eta}^2+I_{\eta}^2}{2}
      \cosh\left(\frac{\Delta\Gamma\,t}{2}\right) 
  + \frac{1-R_{\eta}^2-I_{\eta}^2}{2}
      \cos\left(\Delta m\,t\right)
  \right. \nonumber \\
&& \qquad \qquad \qquad
  \left.
+ \eta_i
  \left[ 
      (R_{\eta}\cos\phi-I_{\eta}\sin\phi)
        \sinh\left(\frac{\Delta\Gamma\,t}{2}\right)
    - (R_{\eta}\sin\phi+I_{\eta}\cos\phi)
        \sin\left(\Delta m\,t\right)
  \right]
  \right \},
\end{eqnarray}
where $\Delta m \equiv m_H-m_L$ and $\Delta \Gamma \equiv
\Gamma_H-\Gamma_L$.

Similarly, one uses the time evolution for the conjugate mode to get,
along with Eq.~(\ref{conjugatedecaymatrix}), for example, the
corresponding time evolution formulas for $|\bar A_{\eta}(t)|^2$:
\begin{eqnarray}
\label{conjugateampsquare}
&& |\bar A_{\eta}(t)|^2 = |A_{\eta}(0)|^2 e^{-\Gamma t}
  \Biggl\{ 
    \frac{1+R_{\eta}^2+I_{\eta}^2}{2}
      \cosh\left(\frac{\Delta\Gamma\,t}{2}\right) 
  - \frac{1-R_{\eta}^2-I_{\eta}^2}{2}
      \cos\left(\Delta m\,t\right)
  \Biggr. \nonumber \\
&& \qquad \qquad \qquad
  \Biggl.
+ \eta_i
  \left[ 
      (R_{\eta}\cos\phi-I_{\eta}\sin\phi)
        \sinh\left(\frac{\Delta\Gamma\,t}{2}\right)
    + (R_{\eta}\sin\phi+I_{\eta}\cos\phi)
        \sin\left(\Delta m\,t\right)
  \right]
  \Biggr\}.
\end{eqnarray}
A complete list of all the observable amplitude bilinears and their CP
conjugates is given in the Appendix.

Before we proceed the discussion, let's define the CP asymmetry
parameters, $\zeta_i(t) \equiv K_i(t) - \bar K_i(t)$ for
$i=1,2,3...,6$ and $\xi_i(t) \equiv L_i(t) - \bar L_i(t)$ for
$i=4,5,6$.  These nine parameters measure the changes of the amplitude
bilinears under the CP transformation.  For instance, from
Eqs.~(\ref{ampsquare}) and (\ref{conjugateampsquare}), we obtain
\begin{equation}
\zeta_1(t) = K_1(0) e^{-\Gamma t}
   \left[
      \left( 1-R_0^2-I_0^2 \right) \cos \left( \Delta m\,t \right)
    -2\left( R_0 \sin \phi + I_0 \cos \phi \right)
         \sin \left( \Delta m\,t \right)
   \right].
\end{equation}
This relation along with others for $K_{2,3}(t)$ provide information
on $\phi$ given $\Delta m$ and $\Gamma$ extracted from other
experiments and theoretical estimates of $K_{1,2,3}(0)$,
$R_{0,\|,\perp}$ and $I_{0,\|,\perp}$.

%%%%%%%%%%%%%%%%%%%%%%%%%%%%%%%%%%%%%%%%%%%%%%%%%%%%%%%%%%%%%%%

\section{Case I: No Time Evolution}
If we take $t=0$ in Eqs.~(\ref{amp})-(\ref{im}) and
(\ref{conjugateamp})-(\ref{conjugateim}), we get the bilinear formulas
for neutral $B$ meson decays at time $t=0$, or the charged $B$ meson
decays.  The relations between the conjugate amplitude bilinears and
amplitude bilinears are
\begin{eqnarray}
\label{notime}
\bar K_i &=& \left( R_{\eta}^2 + I_{\eta}^2 \right) K_i, \,
  {\rm for} \,\, i=1,2,3,
\nonumber \\
\bar K_4 &=& \left( R_{\|}R_0 + I_{\|}I_0 \right) K_4
            - \left( I_{\|}R_0 - R_{\|}I_0 \right) L_4,
\nonumber \\
\bar K_{5,6} &=& \left( R_{\perp}R_{0,\|} + I_{\perp}I_{0,\|} \right) K_{5,6}
            + \left( I_{\perp}R_{0,\|} - R_{\perp}I_{0,\|} \right) L_{5,6},
\\
\bar L_4 &=& \left( R_{\|}R_0 + I_{\|}I_0 \right) L_4
            + \left( I_{\|}R_0 - R_{\|}I_0 \right) K_4,
\nonumber \\
\bar L_{5,6} &=& \left( R_{\perp}R_{0,\|} + I_{\perp}I_{0,\|} \right) L_{5,6}
            - \left( I_{\perp}R_{0,\|} - R_{\perp}I_{0,\|} \right) K_{5,6},
\nonumber
\end{eqnarray}

As discussed in the paragraph after Eq.~(\ref{randi}), if none of the
relative strong and weak phases are trivial, {\it i.e.}, $0$ or $\pi$,
CP asymmetry exists in the above bilinears.  However, if there are no
strong phases (including all the factored strong phases and relative
phases) but the relative weak phase is nontrivial, then one can
simplify the above equations to get $\bar K_{1,2,3,4}=K_{1,2,3,4}$,
$\bar L_{5,6}=L_{5,6}$, $\bar K_{5,6}=-K_{5,6}$, and $\bar L_4=-L_4$.
This effect is purely due to that fact that there is a relative weak
phase and $Im[A_0^* A_{\|}]$, $Im[A_0^* A_{\perp}]$, and $Im[A_{\|}^*
A_{\perp}]$ are CP odd quantities \cite{CW99}.

The observation of CP asymmetries in any of the bilinears indicates
that nontrivial strong and weak phases are involved in the decay.
Therefore, if the relative weak phase within the Standard Model is
trivial, that is, effectively only one weak amplitude dominates, then
no CP asymmetry will be observed among all the bilinears.

The formulae presented in this section can be applied to $B_u^+ \to
D^{*+} \bar D^*$, $B_u^+ \to J/\Psi \rho^+$, $B_u^+ \to D_s^{*+} \bar
D^*$, and $B_u^+ \to J/\Psi K^{*+}$.  One can only measure $K_{1-6}$
in the first decay mode because it is a {\it Type I} decay.  There is
no nontrivial weak phases in the latter two decays.  Therefore, one
should not expect to observe CP asymmetries in the observables; but
$K_{5,6}$ and $L_4$ may be nonzero, and provide evidence for strong
phases due to final state interactions.  The observation of CP
asymmetries in such modes indicates new CP violating source from
physics beyond the Standard Model.

%%%%%%%%%%%%%%%%%%%%%%%%%%%%%%%%%%%%%%%%%%%%%%%%%%%%%%%%%%%%%%%

\section{Case II: No Strong Phases}

If there is no strong phases involved in the decays, then $R_{\eta}^2
+ I_{\eta}^2 = 1$.  One can write
\begin{equation}
\label{alpha}
R_{\eta} = \cos 2\alpha_{\eta}, \qquad
I_{\eta} = -\sin 2 \alpha_{\eta},
\end{equation}
where $\alpha_{\eta}$ is the phase of $T_{\eta} + P_{\eta} e^{i
\phi_w}$.  With Eq.~(\ref{alpha}) and the definition of the phase
$\phi$ in Eq.~(\ref{phi}), one can get the nine CP asymmetry
parameters
\begin{eqnarray}
\zeta_i(t) &=& 
  2 \eta_i \, K_i(0) e^{-\Gamma t} \sin (2\alpha_{\eta} - \phi) 
    \sin (\Delta m t),\, {\rm for}\,\, i=1,2,3, \\
\zeta_4(t) &=&
  K_4(0) e^{-\Gamma t}
    \left[
      \sin (2\alpha_{\|}-\phi) + \sin (2\alpha_0-\phi)
    \right] 
    \sin (\Delta m t) \nonumber \\
&&
- L_4(0) e^{-\Gamma t}
    \biggl[
      \cos (2\alpha_{\|}-\phi) - \cos (2\alpha_0-\phi)
    \biggr] \sin (\Delta m t), \\
\label{zeta5nostrong}
\zeta_5(t) &=&
  K_5(0) e^{-\Gamma t}
    \Biggl\{
      \cosh \left(\frac{\Delta\Gamma t}{2}\right) + \cos (\Delta m t)
\nonumber \\
&& \qquad
    + \biggl[
        \cos (2\alpha_0-\phi) - \cos (2\alpha_{\perp}-\phi)
      \biggr] \sinh \left(\frac{\Delta\Gamma t}{2}\right)
\nonumber \\
&& \qquad
    - \cos (2\alpha_0-2\alpha_{\perp})
      \biggl[
        \cosh \left(\frac{\Delta\Gamma t}{2}\right)
      - \cos (\Delta m t)
      \biggr]
    \Biggr\} \nonumber \\
&&
+ L_5(0) e^{-\Gamma t}
    \Biggl\{
      \biggl[
         \sin (2\alpha_0-\phi) + \sin (2\alpha_{\perp}-\phi)
      \biggr] \sinh \left(\frac{\Delta\Gamma t}{2}\right)
\nonumber \\
&& \qquad
    - \sin (2\alpha_0-2\alpha_{\perp})
      \biggl[
        \cosh \left(\frac{\Delta\Gamma t}{2}\right)
      - \cos (\Delta m t)
      \biggr]
    \Biggl\}, \\
\xi_4(t) &=&
  L_4(0) e^{-\Gamma t}
    \Biggl\{
      \cosh \left(\frac{\Delta\Gamma t}{2}\right) + \cos (\Delta m t)
\nonumber \\
&& \qquad
    + \biggl[
        \cos (2\alpha_0-\phi) + \cos (2\alpha_{\|}-\phi)
      \biggr] \sinh \left(\frac{\Delta\Gamma t}{2}\right)
\nonumber \\
&& \qquad
    + \cos (2\alpha_0-2\alpha_{\|})
      \biggl[
        \cosh \left(\frac{\Delta\Gamma t}{2}\right)
      - \cos (\Delta m t)
      \biggr]
    \Biggr\} \nonumber \\
&&
- K_4(0) e^{-\Gamma t}
    \Biggl\{
      \biggl[
         \sin (2\alpha_{\|}-\phi) - \sin (2\alpha_0-\phi)
      \biggr] \sinh \left(\frac{\Delta\Gamma t}{2}\right)
\nonumber \\
&& \qquad
    - \sin (2\alpha_{\|}-2\alpha_0)
      \biggl[
        \cosh \left(\frac{\Delta\Gamma t}{2}\right)
      - \cos (\Delta m t)
      \biggr]
    \Biggl\}, \\
\label{xi5nostrong}
\xi_5(t) &=&
  L_5(0) e^{-\Gamma t}
    \left[
      \sin (2\alpha_{\bot}-\phi) + \sin (2\alpha_0-\phi)
    \right] 
    \sin (\Delta m t) \nonumber \\
&&
+ K_5(0) e^{-\Gamma t}
    \biggl[
      \cos (2\alpha_{\bot}-\phi) - \cos (2\alpha_0-\phi)
    \biggr] \sin (\Delta m t).
\end{eqnarray}
The formulas for $\zeta_6(t)$ and $\xi_6(6)$ can be obtained by
replacing ``0'' in Eq.~(\ref{zeta5nostrong}) and (\ref{xi5nostrong})
by ``$\|$''.  Thus, in principle, one may extract information about
the phase combinations $2\alpha_{0,\|,\perp}-\phi$.  One should notice
that $\xi_4$ and $\zeta_{5,6}$ can be nonzero at $t=0$, whereas the
others are identically zero.  Although the assumption of no strong
phases is unlikely to be true in charming decays, it may be applied to
charmless decays such as $B \to \rho \rho$.

%%%%%%%%%%%%%%%%%%%%%%%%%%%%%%%%%%%%%%%%%%%%%%%%%%%%%%%%%%%%%%%

\section{Case III: No Relative Weak Phase}

If there is no relative weak phase in each transversity amplitude,
namely, $\phi_w=0$, then one gets $R_{\eta}=1$ and $I_{\eta}=0$.  This
case is equivalent to the cases where only one type of amplitude
dominates the decay process.  For completeness, we list time
evolutions of the nine observables in Tables I and II \footnote{While
we agree with \cite{DDF99} in $K_{1-6}$ and ${\bar K}_{1-4}$, our
${\bar K}_{5,6}$ differ from theirs by an overall minus sign.}.
\begin{table}[ht]
\begin{tabular}{c|c}
Bilinear & Time evolution \\ \hline
$K_i(t)$ & $K_i(0) e^{-\Gamma t}
   \biggl\{
      \cosh \left(\frac{\Delta \Gamma t}{2}\right)
      + \eta_i \left[ \cos \phi \sinh \left(\frac{\Delta \Gamma t}{2}\right)
                   -\sin \phi \sin \left(\Delta m t\right)
             \right]
   \biggr\}$ for $i=1,2,3$, \\
$K_4(t)$ & $K_4(0) e^{-\Gamma t}
   \biggl[
      \cosh \left(\frac{\Delta \Gamma t}{2}\right)
      + \cos \phi \sinh \left(\frac{\Delta \Gamma t}{2}\right)
      - \sin \phi \sin \left(\Delta m t\right)
   \biggr]$ \\
$L_4(t)$ & $L_4(0) e^{-\Gamma t}
   \biggl[
      \cosh \left(\frac{\Delta \Gamma t}{2}\right)
      + \cos \phi \sinh \left(\frac{\Delta \Gamma t}{2}\right)
      - \sin \phi \sin \left(\Delta m t\right)
   \biggr]$ \\
$K_{5,6}(t)$ & 
 $K_{5,6}(0) e^{-\Gamma t} \cos \left( \Delta m t \right)
- L_{5,6}(0) e^{-\Gamma t}
     \biggl[
        \sin \phi \sinh \left(\frac{\Delta \Gamma t}{2}\right)
      + \cos \phi \sin \left(\Delta m t\right)
     \biggr]$ \\
$L_{5,6}(t)$ & 
 $L_{5,6}(0) e^{-\Gamma t} \cos \left( \Delta m t \right)
+ K_{5,6}(0) e^{-\Gamma t}
     \biggl[
        \sin \phi \sinh \left(\frac{\Delta \Gamma t}{2}\right)
      + \cos \phi \sin \left(\Delta m t\right)
     \biggr]$
\end{tabular} \vspace{6pt}
\caption{Time evolutions of observables in the decay of an initially
pure $B_q$ meson into a self-conjugate state of two vector mesons.}
\end{table}
\begin{table}[ht]
\begin{tabular}{c|c}
Bilinear & Time evolution \\ \hline
$\bar K_i(t)$ & $K_i(0) e^{-\Gamma t}
   \biggl\{
      \cosh \left(\frac{\Delta \Gamma t}{2}\right)
      + \eta_i \left[ \cos \phi \sinh \left(\frac{\Delta \Gamma t}{2}\right)
                   +\sin \phi \sin \left(\Delta m t\right)
             \right]
   \biggr\}$ for $i=1,2,3$ \\
$\bar K_4(t)$ & $K_4(0) e^{-\Gamma t}
   \biggl[
      \cosh \left(\frac{\Delta \Gamma t}{2}\right)
      + \cos \phi \sinh \left(\frac{\Delta \Gamma t}{2}\right)
      + \sin \phi \sin \left(\Delta m t\right)
   \biggr]$ \\
$\bar L_4(t)$ & $L_4(0) e^{-\Gamma t}
   \biggl[
      \cosh \left(\frac{\Delta \Gamma t}{2}\right)
      + \cos \phi \sinh \left(\frac{\Delta \Gamma t}{2}\right)
      + \sin \phi \sin \left(\Delta m t\right)
   \biggr]$ \\
$\bar K_{5,6}(t)$ & 
 $K_{5,6}(0) e^{-\Gamma t} \cos \left( \Delta m t \right)
- L_{5,6}(0) e^{-\Gamma t}
     \biggl[
      - \sin \phi \sinh \left(\frac{\Delta \Gamma t}{2}\right)
      + \cos \phi \sin \left(\Delta m t\right)
     \biggr]$ \\
$\bar L_{5,6}(t)$ &
 $L_{5,6}(0) e^{-\Gamma t} \cos \left( \Delta m t \right)
+ K_{5,6}(0) e^{-\Gamma t}
     \biggl[
      - \sin \phi \sinh \left(\frac{\Delta \Gamma t}{2}\right)
      + \cos \phi \sin \left(\Delta m t\right)
     \biggr]$
\end{tabular} \vspace{6pt}
\caption{Time evolutions of observables in the decay of an initially
pure ${\bar B}_q$ meson into a self-conjugate state of two vector
mesons.}
\end{table}

So the nine CP asymmetry parameters are
\begin{eqnarray}
\label{noweaktime}
\zeta_i(t) &=& 
  -2 \eta_i \, K_i(0) e^{-\Gamma t} \sin \phi 
    \sin (\Delta m t),\, {\rm for}\,\, i=1,2,3, \nonumber \\
\zeta_4(t) &=&
  - 2 K_4(0) e^{-\Gamma t} \sin \phi \sin (\Delta m t), \nonumber \\
\zeta_{5,6}(t) &=&
  - 2 L_{5,6}(0) e^{-\Gamma t}
      \sin \phi \sinh \left(\frac{\Delta\Gamma t}{2}\right), \\
\xi_4(t) &=&
  - 2 L_4(0) e^{-\Gamma t} \sin \phi \sin (\Delta m t), \nonumber \\
\xi_{5,6}(t) &=&
    2 K_{5,6}(0) e^{-\Gamma t}
      \sin \phi \sinh \left(\frac{\Delta\Gamma t}{2}\right). \nonumber
\end{eqnarray}
These equations hold even if there are nontrivial strong phases.

If we fix the overall strong phases of the transversity amplitudes by
the following convention: $A_{\perp}(0)=|A_{\perp}(0)|$,
$A_0(0)=|A_0(0)|e^{-i \delta_0}$, and $A_{\|}(0)=|A_{\|}(0)|e^{-i
\delta_{\|}}$ \footnote{Here we ignore the common weak factor that
will be cancelled in all amplitude bilinears.}, then $K_{4,5,6}(0)$
and $L_{4,5,6}(0)$ can be rewritten as
\begin{eqnarray}
& K_4(0) = \sqrt{K_1(0)K_2(0)} \cos(\delta_0-\delta_{\|}), & \qquad
  L_4(0) = \sqrt{K_1(0)K_2(0)} \sin(\delta_0-\delta_{\|}), \nonumber \\
& K_5(0) = \sqrt{K_1(0)K_3(0)} \cos(\delta_0), & \qquad
  L_5(0) = \sqrt{K_1(0)K_3(0)} \sin(\delta_0), \\
& K_6(0) = \sqrt{K_2(0)K_3(0)} \cos(\delta_{\|}), & \qquad
  L_6(0) = \sqrt{K_2(0)K_3(0)} \sin(\delta_{\|}). \nonumber
\end{eqnarray}
One can readily reach four relations among them:
\begin{eqnarray}
\label{consistency}
K_1(0) K_2(0) = K_4(0)^2+L_4(0)^2, & \qquad
K_2(0) K_3(0) = K_6(0)^2+L_6(0)^2, \nonumber\\
K_3(0) K_1(0) = K_5(0)^2+L_5(0)^2, & \qquad
{\displaystyle \frac{L_4(0)}{K_4(0)} = 
\frac{L_5(0) K_6(0) - K_5(0) L_6(0)}{K_5(0) K_6(0) + L_5(0) L_6(0)}}.
\end{eqnarray}
All experimentally measured nine amplitude bilinears should obey the
above consistency relations.  If the strong phases $\delta_0$ and
$\delta_{\|}$ are nontrivial, one could possibly get sizeable
$L_{4,5,6}$ that can be observed experimentally.  We can then obtain
information on the strong phases $\delta_0$, $\delta_{\|}$, the mass
difference $\Delta m$, the decay width difference $\Delta \Gamma$, and
$\sin \phi$ from Eqs.~(\ref{noweaktime}).  Since some of them share
the same time evolution pattern, they also provide a consistency check
for the experimental results.

At $t=0$, there is no $CP$ asymmetry at all.  So for charged $B$
decays where one weak amplitude dominates in the Standard Model, one
should get the same amplitude bilinears for the particle and its
conjugate modes.  However, for neutral $B$ decays, the asymmetries
develop as time goes on due to the mixing effect.  In particular,
$\zeta_{1-4}(t)$ and $\xi_4(t)$ have a sinusoidal time dependence,
while $\zeta_{5,6}$ and $\xi_{5,6}(t)$ decay exponentially at $B_L$'s
decay rate, $\Gamma_L$, in the large $t$ limit.

It is, nevertheless, interesting to look at the time integrated
quantities for sizeable CP or T violating effects.  The particle total
decay rate, after time integration, is
\begin{eqnarray}
\label{intrate}
\int_0^{\infty} dt
    \left[ K_1(t)+K_2(t)+K_3(t) \right]
&=&
\frac1{\Gamma} \biggl\{
  \frac4{4-y^2}
  \left[
     K_1(0)+K_2(0)+K_3(0)
  \right] \biggr. \nonumber \\
&&
\biggl.
\left(\cos \phi \frac{2y}{4-y^2} - \sin \phi \frac{x}{1+x^2} \right)
\left[ K_1(0)+K_2(0)-K_3(0) \right] \biggr\}.
\end{eqnarray}
Similarly, the time integrated anti-particle total decay rate can be
obtained by simply reversing the sign of $\phi$ in
Eq.~(\ref{intrate}).  One can obtain $\sin \phi$ from the asymmetry
between the time integrated total rates of conjugate modes and $\cos
\phi$ from the untagged analysis.  This then eliminates the discrete
ambiguity in the angle $\phi$.

By integrating Eq.~(\ref{noweaktime}) from $t=0$ to $t=\infty$, we
find the asymmetries to be
\begin{eqnarray}
\int_0^{\infty} dt \zeta_i(t) &=& 
  -2 \eta_i \, K_i(0) \frac1{\Gamma} \frac{2x}{1+x^2} \sin \phi,
      \, {\rm for}\,\, i=1,2,3, \nonumber \\
\int_0^{\infty} dt \zeta_4(t) &=&
  - 2 K_4(0) \frac1{\Gamma} \frac{2x}{1+x^2} \sin \phi, \nonumber \\
\int_0^{\infty} dt \zeta_{5,6}(t) &=&
  - 2 L_{5,6}(0) \frac1{\Gamma} \frac{2y}{4-y^2} \sin \phi, \\
\int_0^{\infty} dt \xi_4(t) &=&
  - 2 L_4(0) \frac1{\Gamma} \frac{2x}{1+x^2} \sin \phi, \nonumber \\
\int_0^{\infty} dt \xi_{5,6}(t) &=&
    2 K_{5,6}(0) \frac1{\Gamma} \frac{2y}{4-y^2} \sin \phi, \nonumber
\end{eqnarray}
In the above equations, $x \equiv \Delta m / \Gamma$ and $y \equiv
\Delta \Gamma / \Gamma$.  For $B_d$, $x=0.73$ and $y$ is negligibly
small; for $B_s$, $x>14.0\,({\rm CL}=95\%)$ and $y<0.67\,({\rm
CL}=95\%)$ \cite{PDG98}.  From these relations, one can also directly
extract $\sin \phi$ given the information about the amplitude
bilinears at initial time.

In principle, one can extract information about $\phi$ and strong
phases $\delta_0$, $\delta_{\|}$ either from the time-dependent CP
asymmetries $\zeta$'s and $\xi$'s or from the integrated asymmetries
once the bilinears are determined experimentally or from models.
$\sin 2\beta$ has been measured from the mixing-induced CP asymmetry
of $B_d \to J/\Psi K_S$ \cite{CDF98}.  For $B_d \to J/\Psi K^* (\to
\pi K_S)$, the $B_d-\bar B_d$ and $K-\bar K$ mixings and the CKM
factor in the weak decay amplitude also give $\phi=-2\beta$
\cite{N92,DDF99}.  Therefore, this offers an alternative way of
measuring $\sin 2\beta$ through the angular distribution analysis of
tagged $B_d$ decays.  In addition, the $\cos 2\beta$ dependence in
$K_{5,6}$ and $L_{5,6}$ helps resolving the discrete ambiguity of the
CKM angle $\beta$ \cite{DDF99}.

For $B_s \to D_s^{*+} D_s^{*-}$ and $B_s \to J/\Psi \phi$,
$\phi=2\lambda^2\eta={\cal O}(0.03)$ to an extremely good
approximation, where $\lambda=0.22$ is the Cabibbo angle and $\eta$ is
one of the Wolfenstein parameters \cite{W83}.  In this case, the CP
asymmetries $\zeta_{1,2,3}(t)$ can be used to provide an unambiguous
determination of the sign of $\phi$, and therefore the sign of $\eta$.

The decays that one may apply the results in this section to include:
$B_s \to D_s^{*+} D_s^{*-}$, $B_d \to J/\Psi K^* (\to \pi K_S)$, $B_d
\to J/\Psi \phi$.  The first mode is a {\it Type III} decay, whereas
the latter two are {\it Type II} decays.

Notice that $\sin \phi$ appears in all CP asymmetries, where $\phi$ is
the phase of mixing and the single CKM factor involved in the decay
amplitude.  Since the amplitude has only one CKM factor, no CP
violation effects would be found in the nine observables at $t=0$.
Yet the mixing will produce differences between the particle and
anti-particle decay modes as time goes on.  So any observation of the
CP asymmetries in such modes indicates CP violation due to mixing and
decay.

\section{Numerical Calculation}

In this section, we apply the factorization hypothesis
\cite{FS78,BSW85,BSW87,KG79,B89,BHP96} to the calculation of hadronic
decay amplitudes.  In general, factorization is expected to hold more
strongly for color-allowed processes, such as $B_q \to D_s^{*+} \bar
D_q^*$ with $q \in {s,d,u}$, though it is doubtful in color-suppressed
modes, such as $B_q \to J/\Psi V$ with $(q,V) \in
{(s,\phi),(d,K^{*0}),(u,K^{*+})}$ \cite{AYOPR93,BH92,R90,MRR91,DDF99}.
Throughout the calculations, we ignore the strong phases
$\theta_{\eta}$ in Eq.~(\ref{decaymatrix}) for simplicity but keep the
strong phases in the Wilson coefficients \cite{BSS79}.  These
strong phases may be extracted from experimental data as mentioned in
the previous section.

In our calculations, the Wolfenstein parameters are
$(\rho,\eta)=(0.18,0.37)$.  The decay constants used are $F_{D^*}=230
MeV$, $F_{D_s^*}=275 MeV$, $F_{J/\Psi}=394 MeV$, $F_{K^*}=221 MeV$,
and $F_{\phi}=237 MeV$.  Extracting dominant Wilson coefficients,
$a_1$ and $a_2$, from experimental decay rates has been performed in
\cite{CY99}.  We apply their results to $B^{\pm}$ and $B_d$ decays
with $b \to s$ quark level transitions, i.e., $B_u \to D_s^{*+} \bar
D^{*0}$, $B_d \to D_s^{*+} D^{*-}$, $B_u \to J/\Psi K^{*+}$, and $B_d
\to J/\Psi K^{*0}$.  We then extend the results to other decays
according to the final state configurations, color-allowed ($B_q \to
D_s^{*+} \bar D_q^*$ and $B_q \to D^{*+} D^*$) or color-suppressed
($B_q \to J/\Psi V$), charged or neutral.  Assuming heavy quark
symmetry, we use the $B \to D^*$ decay form factors for the $B \to
D_s^*$ transitions.

The bilinears $K_i$ and $L_i$ in the following tables are normalized
by dividing with $\Gamma_0 \equiv K_1+K_2+K_3$.  The branching ratio
asymmetry is defined by
\begin{equation}
a_{CP} \equiv \frac{{\cal A} - \bar {\cal A}}{{\cal A} + \bar {\cal A}},
\nonumber
\end{equation}
where ${\cal A}$ and $\bar {\cal A}$ are the branching ratios for the
particle and antiparticle decays, respectively.  In the following
tables, we list all nine amplitude bilinears for each mode even if
$L_{4,5,6}$ may not be able to be observed from the angular
distributions of some of them (Type I decays).

In Tables III and IV, we take the modified BSW (or BSW II) model
\cite{BSW85,BSW87} for the form factors in the evaluation of hadronic
matrix elements.  In Tables V and VI, the Neubert-Stech (NS) model
\cite{NS97} is used.  The relativistic light-front (LF) model
\cite{J90,CCW97} is applied to the calculations in Tables VII and
VIII.

We see that in general (1) there are no CP asymmetries for the nine
normalized observables at the initial time (yet the CP asymmetries do
exist for the unnormalized observables); (2) the branching ratio CP
asymmetry is larger in decays involving the $b \to d$ quark level
processes because of the relative phase between the CKM factors of two
weak amplitudes; and (3) $L_4$, $K_5$, and $K_6$ that involve the
imaginary parts of the amplitude bilinears are essentially zero
because the tree amplitudes dominates over the penguin contributions
in these decays and we ignore possible final state interaction phases.

Although the branching ratios and nine observables may vary as one
uses different form factor models, the asymmetries are roughly the
same.  It is found that $b \to d$ type decays have larger asymmetries
than $b \to s$ type ones, as one would expect.  From the models we
analyze, the asymmetries for $B_u^+ \to D^{*+} \bar D^{*0}$ and $B_d
\to D^{*+} D^{*-}$ range from $-3.29\%$ to $-3.53\%$ and from
$-4.14\%$ to $-4.46\%$, respectively.  Unlike the charmless decays
where $A_0$ is the dominant component in the transversity amplitudes,
both $A_0$ and $A_{\|}$ are about the same size.  The parity odd
component $A_{\perp}$ is still small in $B \to D^* D^*$ type
transitions, but larger in $B \to J/\Psi V$ decays.

\begin{table}[vc]
\begin{tabular}{c|cccccccccc}
Processes & Br($\times 10^{-3}$) & $K_1$ & $K_2$ & $K_3$ & $K_4$ &
$K_5$ & $K_6$ & $L_4$ & $L_5$ & $L_6$ \\ \hline
$B_u^+ \to D^{*+} \bar D^{*0}$ &
$1.18$ & $0.514$ & $0.415$ & $0.071$ & $-0.462$ &
$0$ & $0$ & $0$ & $-0.172$ & $0.191$ \\
$B_u^- \to D^{*-} D^{*0}$ &
$1.26$ & $0.514$ & $0.415$ & $0.071$ & $-0.462$ &
$0$ & $0$ & $0$ & $-0.172$ & $0.191$ \\ \hline
$B_u^+ \to J/\Psi \rho^+$ &
$0.0832$ & $0.306$ & $0.413$ & $0.281$ & $-0.356$ &
$0$ & $0$ & $0$ & $-0.340$ & $0.293$ \\
$B_u^- \to J/\Psi \rho^-$ &
$0.0839$ & $0.306$ & $0.413$ & $0.281$ & $-0.356$ &
$0$ & $0$ & $0$ & $-0.340$ & $0.293$ \\ \hline \hline
$B_d^0 \to D^{*+} D^{*-}$ &
$0.778$ & $0.514$ & $0.415$ & $0.071$ & $-0.462$ &
$0$ & $0$ & $0$ & $-0.172$ & $0.191$ \\
$\bar B_d^0 \to D^{*+} D^{*-}$ &
$0.846$ & $0.514$ & $0.415$ & $0.071$ & $-0.462$ &
$0$ & $0$ & $0$ & $-0.172$ & $0.191$ \\ \hline
$B_d^0 \to J/\Psi \rho^0$ &
$0.0832$ & $0.306$ & $0.413$ & $0.281$ & $-0.356$ &
$0$ & $0$ & $0$ & $-0.340$ & $0.293$ \\
$\bar B_d^0 \to J/\Psi \rho^0$ &
$0.0839$ & $0.306$ & $0.413$ & $0.281$ & $-0.356$ &
$0$ & $0$ & $0$ & $-0.340$ & $0.293$ \\ \hline \hline
$B_s^0 \to D_s^{*-} D^{*+}$ &
$1.05$ & $0.514$ & $0.419$ & $0.067$ & $-0.464$ &
$0$ & $0$ & $0$ & $-0.168$ & $0.186$ \\
$\bar B_s^0 \to D_s^{*+} D^{*-}$ &
$1.07$ & $0.514$ & $0.419$ & $0.067$ & $-0.464$ &
$0$ & $0$ & $0$ & $-0.168$ & $0.186$ \\ \hline
$B_s^0 \to J/\Psi K^{*0}$ &
$0.1076$ & $0.354$ & $0.390$ & $0.256$ & $-0.372$ &
$0$ & $0$ & $0$ & $-0.316$ & $0.301$ \\
$\bar B_s^0 \to J/\Psi \bar K^{*0}$ &
$0.1084$ & $0.354$ & $0.390$ & $0.256$ & $-0.372$ &
$0$ & $0$ & $0$ & $-0.316$ & $0.301$ \\
\end{tabular} \vspace{6pt}
\caption{Charming $B \to VV$ decays involving the $b \to d$ underlying
quark processes.  BSW II form factors are used in this table.  The
branching ratio asymmetries of the paired modes are, from top to
bottom, $-3.29\%$, $-0.38\%$, $-4.14\%$, $-0.38\%$, $-0.97\%$, and
$-0.38\%$, respectively.}
\end{table}
\begin{table}[ht]
\begin{tabular}{c|cccccccccc}
Processes & Br($\times 10^{-3}$) & $K_1$ & $K_2$ & $K_3$ & $K_4$ &
$K_5$ & $K_6$ & $L_4$ & $L_5$ & $L_6$ \\ \hline
$B_u^+ \to D_s^{*+} D^{*0}$ &
$34.6$ & $0.491$ & $0.438$ & $0.071$ & $-0.464$ &
$0$ & $0$ & $0$ & $-0.177$ & $0.187$ \\
$B_u^- \to D_s^{*-} \bar D^{*0}$ &
$34.5$ & $0.491$ & $0.438$ & $0.071$ & $-0.464$ &
$0$ & $0$ & $0$ & $-0.177$ & $0.187$ \\ \hline
$B_u^+ \to J/\Psi K^{*+}$ &
$1.988$ & $0.358$ & $0.397$ & $0.245$ & $-0.377$ &
$0$ & $0$ & $0$ & $-0.312$ & $0.296$ \\
$B_u^- \to J/\Psi K^{*-}$ &
$1.987$ & $0.358$ & $0.397$ & $0.245$ & $-0.377$ &
$0$ & $0$ & $0$ & $-0.312$ & $0.296$ \\ \hline \hline
$B_d^0 \to D_s^{*+} D^{*-}$ &
$22.6$ & $0.491$ & $0.438$ & $0.071$ & $-0.464$ &
$0$ & $0$ & $0$ & $-0.177$ & $0.187$ \\
$\bar B_d^0 \to D_s^{*-} D^{*+}$ &
$22.5$ & $0.491$ & $0.438$ & $0.071$ & $-0.464$ &
$0$ & $0$ & $0$ & $-0.177$ & $0.187$ \\ \hline
$B_d^0 \to J/\Psi K^{*0}$ &
$1.988$ & $0.358$ & $0.397$ & $0.245$ & $-0.377$ &
$0$ & $0$ & $0$ & $-0.312$ & $0.296$ \\
$\bar B_d^0 \to J/\Psi \bar K^{*0}$ &
$1.987$ & $0.358$ & $0.397$ & $0.245$ & $-0.377$ &
$0$ & $0$ & $0$ & $-0.312$ & $0.296$ \\ \hline \hline
$B_s^0 \to D_s^{*+} D_s^{*-}$ &
$31.28$ & $0.491$ & $0.442$ & $0.068$ & $-0.466$ &
$0$ & $0$ & $0$ & $-0.173$ & $0.182$ \\
$\bar B_s^0 \to D_s^{*+} D_s^{*-}$ &
$31.25$ & $0.491$ & $0.442$ & $0.068$ & $-0.466$ &
$0$ & $0$ & $0$ & $-0.173$ & $0.182$ \\ \hline
$B_s^0 \to J/\Psi \phi$ &
$2.049$ & $0.353$ & $0.413$ & $0.235$ & $-0.382$ &
$0$ & $0$ & $0$ & $-0.311$ & $0.288$ \\
$\bar B_s^0 \to J/\Psi \phi$ &
$2.049$ & $0.353$ & $0.413$ & $0.235$ & $-0.382$ &
$0$ & $0$ & $0$ & $-0.311$ & $0.288$ \\
\end{tabular} \vspace{6pt}
\caption{Charming $B \to VV$ decays involving the $b \to s$ underlying
quark processes.  BSW II form factors are used in this table.  The
branching ratio asymmetries of the paired modes are, from top to
bottom, $0.18\%$, $0.02\%$, $0.23\%$, $0.02\%$, $0.05\%$, and $0$,
respectively.}
\end{table}
\begin{table}[ht]
\begin{tabular}{c|cccccccccc}
Processes & Br & $K_1$ & $K_2$ & $K_3$ & $K_4$ & $K_5$ & $K_6$ &
$L_4$ & $L_5$ & $L_6$ \\ \hline
$B_u^+ \to D^{*+} \bar D^{*0}$ &
$1.24$ & $0.554$ & $0.396$ & $0.050$ & $-0.469$ &
$0$ & $0$ & $0$ & $-0.140$ & $0.166$ \\
$B_u^- \to D^{*-} D^{*0}$ &
$1.32$ & $0.554$ & $0.396$ & $0.050$ & $-0.469$ &
$0$ & $0$ & $0$ & $-0.140$ & $0.166$ \\ \hline
$B_u^+ \to J/\Psi \rho^+$ &
$0.0957$ & $0.493$ & $0.367$ & $0.141$ & $-0.425$ &
$0$ & $0$ & $0$ & $-0.227$ & $0.263$ \\
$B_u^- \to J/\Psi \rho^-$ &
$0.0963$ & $0.493$ & $0.367$ & $0.141$ & $-0.425$ &
$0$ & $0$ & $0$ & $-0.227$ & $0.263$ \\ \hline \hline
$B_d^0 \to D^{*+} D^{*-}$ &
$0.81$ & $0.554$ & $0.396$ & $0.050$ & $-0.469$ &
$0$ & $0$ & $0$ & $-0.140$ & $0.166$ \\
$\bar B_d^0 \to D^{*+} D^{*-}$ &
$0.88$ & $0.554$ & $0.396$ & $0.050$ & $-0.469$ &
$0$ & $0$ & $0$ & $-0.140$ & $0.166$ \\ \hline
$B_d^0 \to J/\Psi \rho^0$ &
$0.0957$ & $0.493$ & $0.367$ & $0.141$ & $-0.425$ &
$0$ & $0$ & $0$ & $-0.227$ & $0.263$ \\
$\bar B_d^0 \to J/\Psi \rho^0$ &
$0.0963$ & $0.493$ & $0.367$ & $0.141$ & $-0.425$ &
$0$ & $0$ & $0$ & $-0.227$ & $0.263$ \\ \hline \hline
$B_s^0 \to D_s^{*-} D^{*+}$ &
$1.09$ & $0.552$ & $0.401$ & $0.047$ & $-0.470$ &
$0$ & $0$ & $0$ & $-0.137$ & $0.161$ \\
$\bar B_s^0 \to D_s^{*+} D^{*-}$ &
$1.11$ & $0.552$ & $0.401$ & $0.047$ & $-0.470$ &
$0$ & $0$ & $0$ & $-0.137$ & $0.161$ \\ \hline
$B_s^0 \to J/\Psi K^{*0}$ &
$0.131$ & $0.489$ & $0.383$ & $0.128$ & $-0.433$ &
$0$ & $0$ & $0$ & $-0.222$ & $0.250$ \\
$\bar B_s^0 \to J/\Psi \bar K^{*0}$ &
$0.132$ & $0.489$ & $0.383$ & $0.128$ & $-0.433$ &
$0$ & $0$ & $0$ & $-0.222$ & $0.250$ \\
\end{tabular} \vspace{6pt}
\caption{Charming $B \to VV$ decays involving the $b \to d$ underlying
quark processes.  The NS model form factors are used in this table.
The branching ratio asymmetries of the paired modes are, from top to
bottom, $-3.29\%$, $-0.30\%$, $-4.14\%$, $-0.30\%$, $-0.97\%$, and
$-0.30\%$, respectively.}
\end{table}
\begin{table}[ht]
\begin{tabular}{c|cccccccccc}
Processes & Br & $K_1$ & $K_2$ & $K_3$ & $K_4$ & $K_5$ & $K_6$ &
$L_4$ & $L_5$ & $L_6$ \\ \hline
$B_u^+ \to D_s^{*+} D^{*0}$ &
$36.09$ & $0.531$ & $0.420$ & $0.049$ & $-0.472$ &
$0$ & $0$ & $0$ & $-0.144$ & $0.162$ \\
$B_u^- \to D_s^{*-} \bar D^{*0}$ &
$35.96$ & $0.531$ & $0.420$ & $0.049$ & $-0.472$ &
$0$ & $0$ & $0$ & $-0.144$ & $0.162$ \\ \hline
$B_u^+ \to J/\Psi K^{*+}$ &
$2.411$ & $0.484$ & $0.392$ & $0.124$ & $-0.436$ &
$0$ & $0$ & $0$ & $-0.220$ & $0.245$ \\
$B_u^- \to J/\Psi K^{*-}$ &
$2.410$ & $0.484$ & $0.392$ & $0.124$ & $-0.436$ &
$0$ & $0$ & $0$ & $-0.220$ & $0.245$ \\\hline \hline
$B_d^0 \to D_s^{*+} D^{*-}$ &
$23.59$ & $0.531$ & $0.420$ & $0.049$ & $-0.472$ &
$0$ & $0$ & $0$ & $-0.144$ & $0.162$ \\
$\bar B_d^0 \to D_s^{*-} D^{*+}$ &
$23.48$ & $0.531$ & $0.420$ & $0.049$ & $-0.472$ &
$0$ & $0$ & $0$ & $-0.144$ & $0.162$ \\ \hline
$B_d^0 \to J/\Psi K^{*0}$ &
$2.4114$ & $0.484$ & $0.392$ & $0.124$ & $-0.436$ &
$0$ & $0$ & $0$ & $-0.220$ & $0.245$ \\
$\bar B_d^0 \to J/\Psi \bar K^{*0}$ &
$2.4107$ & $0.484$ & $0.392$ & $0.124$ & $-0.436$ &
$0$ & $0$ & $0$ & $-0.220$ & $0.245$ \\ \hline \hline
$B_s^0 \to D_s^{*+} D_s^{*-}$ &
$32.54$ & $0.529$ & $0.425$ & $0.047$ & $-0.474$ &
$0$ & $0$ & $0$ & $-0.141$ & $0.157$ \\
$\bar B_s^0 \to D_s^{*+} D_s^{*-}$ &
$32.51$ & $0.529$ & $0.425$ & $0.047$ & $-0.474$ &
$0$ & $0$ & $0$ & $-0.141$ & $0.157$ \\ \hline
$B_s^0 \to J/\Psi \phi$ &
$3.166$ & $0.478$ & $0.408$ & $0.114$ & $-0.441$ &
$0$ & $0$ & $0$ & $-0.216$ & $0.234$ \\
$\bar B_s^0 \to J/\Psi \phi$ &
$3.165$ & $0.478$ & $0.408$ & $0.114$ & $-0.441$ &
$0$ & $0$ & $0$ & $-0.216$ & $0.234$ \\
\end{tabular} \vspace{6pt}
\caption{Charming $B \to VV$ decays involving the $b \to s$ underlying
quark processes.  The NS model form factors are used in this table.
The branching ratio asymmetries of the paired modes are, from top to
bottom, $0.18\%$, $0.02\%$, $0.23\%$, $0.02\%$, $0.05\%$, and
$0.02\%$, respectively.}
\end{table}
\begin{table}[ht]
\begin{tabular}{c|cccccccccc}
Processes & Br & $K_1$ & $K_2$ & $K_3$ & $K_4$ & $K_5$ & $K_6$ &
$L_4$ & $L_5$ & $L_6$ \\ \hline
$B_u^+ \to D^{*+} \bar D^{*0}$ &
$1.23$ & $0.557$ & $0.376$ & $0.066$ & $-0.458$ &
$0$ & $0$ & $0$ & $-0.158$ & $0.192$ \\
$B_u^- \to D^{*-} D^{*0}$ &
$1.32$ & $0.557$ & $0.376$ & $0.066$ & $-0.458$ &
$0$ & $0$ & $0$ & $-0.158$ & $0.192$ \\ \hline
$B_u^+ \to J/\Psi \rho^+$ &
$0.084$ & $0.592$ & $0.334$ & $0.074$ & $-0.445$ &
$0$ & $0$ & $0$ & $-0.157$ & $0.209$ \\
$B_u^- \to J/\Psi \rho^-$ &
$0.085$ & $0.592$ & $0.334$ & $0.074$ & $-0.445$ &
$0$ & $0$ & $0$ & $-0.157$ & $0.209$ \\ \hline \hline
$B_d^0 \to D^{*+} D^{*-}$ &
$0.81$ & $0.558$ & $0.376$ & $0.066$ & $-0.458$ &
$0$ & $0$ & $0$ & $-0.158$ & $0.192$ \\
$\bar B_d^0 \to D^{*+} D^{*-}$ &
$0.88$ & $0.558$ & $0.376$ & $0.066$ & $-0.458$ &
$0$ & $0$ & $0$ & $-0.158$ & $0.192$ \\ \hline
$B_d^0 \to J/\Psi \rho^0$ &
$0.084$ & $0.592$ & $0.334$ & $0.074$ & $-0.445$ &
$0$ & $0$ & $0$ & $-0.157$ & $0.209$ \\
$\bar B_d^0 \to J/\Psi \rho^0$ &
$0.085$ & $0.592$ & $0.334$ & $0.074$ & $-0.445$ &
$0$ & $0$ & $0$ & $-0.157$ & $0.209$ \\ \hline \hline
$B_s^0 \to D_s^{*-} D^{*+}$ &
$1.10$ & $0.555$ & $0.382$ & $0.063$ & $-0.460$ &
$0$ & $0$ & $0$ & $-0.155$ & $0.187$ \\
$\bar B_s^0 \to D_s^{*+} D^{*-}$ &
$1.19$ & $0.555$ & $0.382$ & $0.063$ & $-0.460$ &
$0$ & $0$ & $0$ & $-0.155$ & $0.187$ \\ \hline
$B_s^0 \to J/\Psi K^{*0}$ &
$0.131$ & $0.536$ & $0.371$ & $0.093$ & $-0.446$ &
$0$ & $0$ & $0$ & $-0.186$ & $0.224$ \\
$\bar B_s^0 \to J/\Psi \bar K^{*0}$ &
$0.132$ & $0.536$ & $0.371$ & $0.093$ & $-0.446$ &
$0$ & $0$ & $0$ & $-0.186$ & $0.224$ \\
\end{tabular} \vspace{6pt}
\caption{Charming $B \to VV$ decays involving the $b \to d$ underlying
quark processes.  The LF model form factors are used in this table.
The branching ratio asymmetries of the paired modes are, from top to
bottom, $-3.53\%$, $-0.29\%$, $-4.46\%$, $-0.29\%$, $-1.03\%$, and
$-0.29\%$, respectively.}
\end{table}
\begin{table}[ht]
\begin{tabular}{c|cccccccccc}
Processes & Br & $K_1$ & $K_2$ & $K_3$ & $K_4$ & $K_5$ & $K_6$ &
$L_4$ & $L_5$ & $L_6$ \\ \hline
$B_u^+ \to D_s^{*+} D^{*0}$ &
$35.42$ & $0.533$ & $0.401$ & $0.067$ & $-0.462$ &
$0$ & $0$ & $0$ & $-0.163$ & $0.188$ \\
$B_u^- \to D_s^{*-} \bar D^{*0}$ &
$35.29$ & $0.533$ & $0.401$ & $0.067$ & $-0.462$ &
$0$ & $0$ & $0$ & $-0.163$ & $0.188$ \\ \hline
$B_u^+ \to J/\Psi K^{*+}$ &
$2.402$ & $0.529$ & $0.381$ & $0.090$ & $-0.449$ &
$0$ & $0$ & $0$ & $-0.185$ & $0.218$ \\
$B_u^- \to J/\Psi K^{*-}$ &
$2.401$ & $0.529$ & $0.381$ & $0.090$ & $-0.449$ &
$0$ & $0$ & $0$ & $-0.185$ & $0.218$ \\ \hline \hline
$B_d^0 \to D_s^{*+} D^{*-}$ &
$23.06$ & $0.533$ & $0.401$ & $0.067$ & $-0.462$ &
$0$ & $0$ & $0$ & $-0.163$ & $0.188$ \\
$\bar B_d^0 \to D_s^{*-} D^{*+}$ &
$22.94$ & $0.533$ & $0.401$ & $0.067$ & $-0.462$ &
$0$ & $0$ & $0$ & $-0.163$ & $0.188$ \\ \hline
$B_d^0 \to J/\Psi K^{*0}$ &
$2.4025$ & $0.529$ & $0.381$ & $0.090$ & $-0.449$ &
$0$ & $0$ & $0$ & $-0.185$ & $0.218$ \\
$\bar B_d^0 \to J/\Psi \bar K^{*0}$ &
$2.4017$ & $0.529$ & $0.381$ & $0.090$ & $-0.449$ &
$0$ & $0$ & $0$ & $-0.185$ & $0.218$ \\ \hline \hline
$B_s^0 \to D_s^{*+} D_s^{*-}$ &
$32.30$ & $0.530$ & $0.407$ & $0.063$ & $-0.464$ &
$0$ & $0$ & $0$ & $-0.160$ & $0.183$ \\
$\bar B_s^0 \to D_s^{*+} D_s^{*-}$ &
$32.27$ & $0.530$ & $0.407$ & $0.063$ & $-0.464$ &
$0$ & $0$ & $0$ & $-0.160$ & $0.183$ \\ \hline
$B_s^0 \to J/\Psi \phi$ &
$2.330$ & $0.496$ & $0.376$ & $0.128$ & $-0.432$ &
$0$ & $0$ & $0$ & $-0.220$ & $0.252$ \\
$\bar B_s^0 \to J/\Psi \phi$ &
$2.329$ & $0.496$ & $0.376$ & $0.128$ & $-0.432$ &
$0$ & $0$ & $0$ & $-0.220$ & $0.252$ \\
\end{tabular} \vspace{6pt}
\caption{Charming $B \to VV$ decays involving the $b \to s$ underlying
quark processes.  The LF model form factors are used in this table.
The branching ratio asymmetries of the paired modes are, from top to
bottom, $0.19\%$, $0.02\%$, $0.25\%$, $0.02\%$, $0.05\%$, and
$0.02\%$, respectively.}
\end{table}
%
%%%%%%%%%%%%%%%%%%%%%%%%%%%%%%%%%%%%%%%%%%%%%%%%%%%%%%%%%%%%%%%

\section{Summary}

The decays of a $B$ meson into two vector mesons, which subsequently
decay into two lighter particles via CP conserving currents, have a
specific pattern in the differential angular distributions.  The
coefficient of each angular function in the distribution is a bilinear
of amplitudes with certain CP properties.  The knowledge of these
amplitude bilinears can help us observe and understand CP violating
effects.  In particular, the time evolution of these bilinears in the
cases of neutral $B$ meson decays further reveals the information such
as the mass and decay width differences and CKM parameters.  In
situations where we can measure the polarization of the final product
particles, all the nine combinations of amplitude bilinears are
observable.

Under certain special circumstances, one can find simple relations
among the nine observables.  Therefore, experimental determination of
them is valuable in testing our theoretical assumptions in the
calculations, such as the factorization hypothesis and form factor
models.  The results are particularly simplified when there is only
one weak amplitude dominating in the decay process.  In such cases,
one can test the Standard Model from the CP asymmetries at $t=0$.  The
time development of CP asymmetries provides a window for observing CP
violations due to mixing effects.

We provide numerical estimates of the observables in 12 sets of
charming $B \to VV$ decays using three different form factor models.
We find that the results do not depend strongly on the models used.
In particular, we find bigger branching ratio asymmetries in $b \to d$
type decays, and those for $B_u^+ \to D^{*+} \bar D^{*0}$ and $B_d \to
D^{*+} D^{*-}$ are as large as $-3\%$ and $-4\%$, respectively.

%%%%%%%%%%%%%%%%%%%%%%%%%%%%%%%%%%%%%%%%%%%%%%%%%%%%%%%%%%%%%%%

{\bf ACKNOWLEDGMENT}

This research work is supported by the Department of Energy under
Grant No. DE-FG02-91ER40682.  The author is grateful to F. Gilman,
L. Wolfenstein for useful discussion and to A. Leibovich for his
comments.  He also would like to thank the hospitality of Academia
Sinica in Taiwan where part of this work is done.

%%%%%%%%%%%%%%%%%%%%%%%%%%%%%%%%%%%%%%%%%%%%%%%%%%%%%%%%%%%%%%%

\begin{appendix}
\section*{Time-Dependent Amplitude Bilinears}

%%%%% Amplitude Bilinears

The time evolutions of the amplitude bilinears are as follows:
\begin{eqnarray}
\label{amp}
&& |A_{\eta}(t)|^2 = |A_{\eta}(0)|^2 e^{-\Gamma t}
  \left \{ 
    \frac{1+R_{\eta}^2+I_{\eta}^2}{2}
      \cosh\left(\frac{\Delta\Gamma\,t}{2}\right) 
  + \frac{1-R_{\eta}^2-I_{\eta}^2}{2}
      \cos\left(\Delta m\,t\right)
  \right. \nonumber \\
&& \qquad \qquad \qquad
  \left.
+ \eta_i
  \left[ 
      (R_{\eta}\cos\phi-I_{\eta}\sin\phi)
        \sinh\left(\frac{\Delta\Gamma\,t}{2}\right)
    - (R_{\eta}\sin\phi+I_{\eta}\cos\phi)
        \sin\left(\Delta m\,t\right)
  \right]
  \right \}, \nonumber \\
&&\;\; {\rm where}\; i=1,2,3\; {\rm for}\; \eta=0,\|,\perp,\;
{\rm respectively, and}\; \eta_{1,2}=-\eta_3=1; \\
&& Re\left[A_0^*(t)A_{\|}(t)\right] =
Re\left[A_0^*(0)A_{\|}(0)\right] \frac{e^{-\Gamma t}}{2} \times
  \nonumber \\
&& \qquad \qquad \qquad
  \Biggl\{
    \left[ \cosh\left(\frac{\Delta\Gamma\,t}{2}\right)
            + \cos\left(\Delta m\,t\right)
    \right]
  \Biggr. \nonumber \\
&& \qquad \qquad \qquad \qquad
  + (R_{\|}\cos\phi-I_{\|}\sin\phi)
        \sinh\left(\frac{\Delta\Gamma\,t}{2}\right)
  - (R_{\|}\sin\phi+I_{\|}\cos\phi)
        \sin\left(\Delta m\,t\right)
  \nonumber \\
&& \qquad \qquad \qquad \qquad
  + (R_0\cos\phi-I_0\sin\phi)
        \sinh\left(\frac{\Delta\Gamma\,t}{2}\right)
  - (R_0\sin\phi+I_0\cos\phi)
        \sin\left(\Delta m\,t\right)
  \nonumber \\
&& \qquad \qquad \qquad \qquad
  \Biggl.
  + (R_{\|}R_0+I_{\|}I_0)
    \left[ \cosh\left(\frac{\Delta\Gamma\,t}{2}\right)
         - \cos\left(\Delta m\,t\right)
    \right]
  \Biggr\} \nonumber \\
&& \qquad \qquad \qquad
-Im\left[A_0^*(0)A_{\|}(0)\right] \frac{e^{-\Gamma t}}{2} \times
  \nonumber \\
&& \qquad \qquad \qquad \qquad
  \Biggl\{
    (R_{\|}\sin\phi+I_{\|}\cos\phi)
        \sinh\left(\frac{\Delta\Gamma\,t}{2}\right)
  + (R_{\|}\cos\phi-I_{\|}\sin\phi)
        \sin\left(\Delta m\,t\right)
  \Biggr. \nonumber \\
&& \qquad \qquad \qquad \qquad
  - (R_0\sin\phi+I_0\cos\phi)
        \sinh\left(\frac{\Delta\Gamma\,t}{2}\right)
  - (R_0\cos\phi-I_0\sin\phi)
        \sin\left(\Delta m\,t\right)
  \nonumber \\
&& \qquad \qquad \qquad \qquad
  \Biggl.
  + (I_{\|}R_0-R_{\|}I_0)
    \left[ \cosh\left(\frac{\Delta\Gamma\,t}{2}\right)
         - \cos\left(\Delta m\,t\right)
    \right]
  \Biggr\}; \\
&& Im\left[A_0^*(t)A_{\|}(t)\right] =
Im\left[A_0^*(0)A_{\|}(0)\right] \frac{e^{-\Gamma t}}{2} \times
  \nonumber \\
&& \qquad \qquad \qquad
  \Biggl\{
    \left[ \cosh\left(\frac{\Delta\Gamma\,t}{2}\right)
            + \cos\left(\Delta m\,t\right)
    \right]
  \Biggr. \nonumber \\
&& \qquad \qquad \qquad \qquad
  + (R_{\|}\cos\phi-I_{\|}\sin\phi)
        \sinh\left(\frac{\Delta\Gamma\,t}{2}\right)
  - (R_{\|}\sin\phi+I_{\|}\cos\phi)
        \sin\left(\Delta m\,t\right)
  \nonumber \\
&& \qquad \qquad \qquad \qquad
  + (R_0\cos\phi-I_0\sin\phi)
        \sinh\left(\frac{\Delta\Gamma\,t}{2}\right)
  - (R_0\sin\phi+I_0\cos\phi)
        \sin\left(\Delta m\,t\right)
  \nonumber \\
&& \qquad \qquad \qquad \qquad
  \Biggl.
  + (R_{\|}R_0+I_{\|}I_0)
    \left[ \cosh\left(\frac{\Delta\Gamma\,t}{2}\right)
         - \cos\left(\Delta m\,t\right)
    \right]
  \Biggr\} \nonumber \\
&& \qquad \qquad \qquad
+Re\left[A_0^*(0)A_{\|}(0)\right] \frac{e^{-\Gamma t}}{2} \times
  \nonumber \\
&& \qquad \qquad \qquad \qquad
  \Biggl\{
    (R_{\|}\sin\phi+I_{\|}\cos\phi)
        \sinh\left(\frac{\Delta\Gamma\,t}{2}\right)
  + (R_{\|}\cos\phi-I_{\|}\sin\phi)
        \sin\left(\Delta m\,t\right)
  \Biggr. \nonumber \\
&& \qquad \qquad \qquad \qquad
  - (R_0\sin\phi+I_0\cos\phi)
        \sinh\left(\frac{\Delta\Gamma\,t}{2}\right)
  - (R_0\cos\phi-I_0\sin\phi)
        \sin\left(\Delta m\,t\right)
  \nonumber \\
&& \qquad \qquad \qquad \qquad
  \Biggl.
  + (I_{\|}R_0-R_{\|}I_0)
    \left[ \cosh\left(\frac{\Delta\Gamma\,t}{2}\right)
         - \cos\left(\Delta m\,t\right)
    \right]
  \Biggr\}; \\
\label{re}
&& Re\left[A_0^*(t)A_{\bot}(t)\right] =
Re\left[A_0^*(0)A_{\bot}(0)\right] \frac{e^{-\Gamma t}}{2} \times
  \nonumber \\
&& \qquad \qquad \qquad
  \Biggl\{
    \left[ \cosh\left(\frac{\Delta\Gamma\,t}{2}\right)
            + \cos\left(\Delta m\,t\right)
    \right]
  \Biggr. \nonumber \\
&& \qquad \qquad \qquad \qquad
  - (R_{\bot}\cos\phi-I_{\bot}\sin\phi)
        \sinh\left(\frac{\Delta\Gamma\,t}{2}\right)
  + (R_{\bot}\sin\phi+I_{\bot}\cos\phi)
        \sin\left(\Delta m\,t\right)
  \nonumber \\
&& \qquad \qquad \qquad \qquad
  + (R_0\cos\phi-I_0\sin\phi)
        \sinh\left(\frac{\Delta\Gamma\,t}{2}\right)
  - (R_0\sin\phi+I_0\cos\phi)
        \sin\left(\Delta m\,t\right)
  \nonumber \\
&& \qquad \qquad \qquad \qquad
  \Biggl.
  - (R_{\bot}R_0+I_{\bot}I_0)
    \left[ \cosh\left(\frac{\Delta\Gamma\,t}{2}\right)
         - \cos\left(\Delta m\,t\right)
    \right]
  \Biggr\} \nonumber \\
&& \qquad \qquad \qquad
+Im\left[A_0^*(0)A_{\bot}(0)\right] \frac{e^{-\Gamma t}}{2} \times
  \nonumber \\
&& \qquad \qquad \qquad \qquad
  \Biggl\{
    (R_{\bot}\sin\phi+I_{\bot}\cos\phi)
        \sinh\left(\frac{\Delta\Gamma\,t}{2}\right)
  + (R_{\bot}\cos\phi-I_{\bot}\sin\phi)
        \sin\left(\Delta m\,t\right)
  \Biggr. \nonumber \\
&& \qquad \qquad \qquad \qquad
  + (R_0\sin\phi+I_0\cos\phi)
        \sinh\left(\frac{\Delta\Gamma\,t}{2}\right)
  + (R_0\cos\phi-I_0\sin\phi)
        \sin\left(\Delta m\,t\right)
  \nonumber \\
&& \qquad \qquad \qquad \qquad
  \Biggl.
  + (I_{\bot}R_0-R_{\bot}I_0)
    \left[ \cosh\left(\frac{\Delta\Gamma\,t}{2}\right)
         - \cos\left(\Delta m\,t\right)
    \right]
  \Biggr\}; \\
\label{im}
&& Im\left[A_0^*(t)A_{\bot}(t)\right] =
Im\left[A_0^*(0)A_{\bot}(0)\right] \frac{e^{-\Gamma t}}{2} \times
  \nonumber \\
&& \qquad \qquad \qquad
  \Biggl\{
    \left[ \cosh\left(\frac{\Delta\Gamma\,t}{2}\right)
            + \cos\left(\Delta m\,t\right)
    \right]
  \Biggr. \nonumber \\
&& \qquad \qquad \qquad \qquad
  - (R_{\bot}\cos\phi-I_{\bot}\sin\phi)
        \sinh\left(\frac{\Delta\Gamma\,t}{2}\right)
  + (R_{\bot}\sin\phi+I_{\bot}\cos\phi)
        \sin\left(\Delta m\,t\right)
  \nonumber \\
&& \qquad \qquad \qquad \qquad
  + (R_0\cos\phi-I_0\sin\phi)
        \sinh\left(\frac{\Delta\Gamma\,t}{2}\right)
  - (R_0\sin\phi+I_0\cos\phi)
        \sin\left(\Delta m\,t\right)
  \nonumber \\
&& \qquad \qquad \qquad \qquad
  \Biggl.
  - (R_{\bot}R_0+I_{\bot}I_0)
    \left[ \cosh\left(\frac{\Delta\Gamma\,t}{2}\right)
         - \cos\left(\Delta m\,t\right)
    \right]
  \Biggr\} \nonumber \\
&& \qquad \qquad \qquad
-Re\left[A_0^*(0)A_{\bot}(0)\right] \frac{e^{-\Gamma t}}{2} \times
  \nonumber \\
&& \qquad \qquad \qquad \qquad
  \Biggl\{
    (R_{\bot}\sin\phi+I_{\bot}\cos\phi)
        \sinh\left(\frac{\Delta\Gamma\,t}{2}\right)
  + (R_{\bot}\cos\phi-I_{\bot}\sin\phi)
        \sin\left(\Delta m\,t\right)
  \Biggr. \nonumber \\
&& \qquad \qquad \qquad \qquad
  + (R_0\sin\phi+I_0\cos\phi)
        \sinh\left(\frac{\Delta\Gamma\,t}{2}\right)
  + (R_0\cos\phi-I_0\sin\phi)
        \sin\left(\Delta m\,t\right)
  \nonumber \\
&& \qquad \qquad \qquad \qquad
  \Biggl.
  + (I_{\bot}R_0-R_{\bot}I_0)
    \left[ \cosh\left(\frac{\Delta\Gamma\,t}{2}\right)
         - \cos\left(\Delta m\,t\right)
    \right]
  \Biggr\}.
\end{eqnarray}
Similar formulas for $Re\left[A_{\|}^*(0)A_{\bot}(0)\right]$ and
$Im\left[A_{\|}^*(0)A_{\bot}(0)\right]$ can be obtained from
Eq.~(\ref{re}) and (\ref{im}) by replacing ``$0$'' with ``$\|$'',
respectively.

%%%%% CP-Conjugate Amplitude Bilinears

The time evolution formulas for the CP conjugate amplitude bilinears
are:
\begin{eqnarray}
\label{conjugateamp}
&& |\bar A_{\eta}(t)|^2 = |A_{\eta}(0)|^2 e^{-\Gamma t}
  \Biggl\{ 
    \frac{1+R_{\eta}^2+I_{\eta}^2}{2}
      \cosh\left(\frac{\Delta\Gamma\,t}{2}\right) 
  - \frac{1-R_{\eta}^2-I_{\eta}^2}{2}
      \cos\left(\Delta m\,t\right)
  \Biggr. \nonumber \\
&& \qquad \qquad \qquad
  \Biggl.
+ \eta_i
  \left[ 
      (R_{\eta}\cos\phi-I_{\eta}\sin\phi)
        \sinh\left(\frac{\Delta\Gamma\,t}{2}\right)
    + (R_{\eta}\sin\phi+I_{\eta}\cos\phi)
        \sin\left(\Delta m\,t\right)
  \right]
  \Biggr\}; \\
&& Re\left[\bar A_0^*(t) \bar A_{\|}(t)\right] =
\biggl[ 
   Re\left[A_0^*(0)A_{\|}(0)\right] \left( R_{\|}R_0+I_{\|}I_0 \right)
 - Im\left[A_0^*(0)A_{\|}(0)\right] \left( I_{\|}R_0-R_{\|}I_0 \right)
\biggr]
  \frac{e^{-\Gamma t}}{2} \times
  \nonumber \\
&& \qquad \qquad \qquad
  \Biggl\{
    \left[ \cosh\left(\frac{\Delta\Gamma\,t}{2}\right)
         + \cos\left(\Delta m\,t\right)
    \right]
  \Biggr. \nonumber \\
&& \qquad \qquad \qquad \qquad
  + \frac1{R_{\|}^2+I_{\|}^2}
    \left[
      (R_{\|}\cos\phi-I_{\|}\sin\phi)
        \sinh\left(\frac{\Delta\Gamma\,t}{2}\right)
    + (R_{\|}\sin\phi+I_{\|}\cos\phi)
        \sin\left(\Delta m\,t\right)
    \right]
  \nonumber \\
&& \qquad \qquad \qquad \qquad
  + \frac1{R_0^2+I_0^2}
    \left[
      (R_0\cos\phi-I_0\sin\phi)
        \sinh\left(\frac{\Delta\Gamma\,t}{2}\right)
    - (R_0\sin\phi+I_0\cos\phi)
        \sin\left(\Delta m\,t\right)
    \right]
  \nonumber \\
&& \qquad \qquad \qquad \qquad
  \Biggl.
  + \frac1{\left(R_{\|}^2+I_{\|}^2\right)\left(R_0^2+I_0^2\right)}
    \left(R_{\|}R_0+I_{\|}I_0\right)
    \left[ \cosh\left(\frac{\Delta\Gamma\,t}{2}\right)
         - \cos\left(\Delta m\,t\right)
    \right]
  \Biggr\} \nonumber \\
&& \qquad \qquad \qquad
- \biggl[ 
    Re\left[A_0^*(0)A_{\|}(0)\right] \left( I_{\|}R_0-R_{\|}I_0 \right)
  + Im\left[A_0^*(0)A_{\|}(0)\right] \left( R_{\|}R_0+I_{\|}I_0 \right)
  \biggr]
  \frac{e^{-\Gamma t}}{2} \times
  \nonumber \\
&& \qquad \qquad \qquad \qquad
  \Biggl\{
  \frac1{R_{\|}^2+I_{\|}^2}
    \left[
    - (R_{\|}\sin\phi+I_{\|}\cos\phi)
        \sinh\left(\frac{\Delta\Gamma\,t}{2}\right)
    + (R_{\|}\cos\phi-I_{\|}\sin\phi)
        \sin\left(\Delta m\,t\right)
    \right]
  \Biggr. \nonumber \\
&& \qquad \qquad \qquad \qquad
  - \frac1{R_0^2+I_0^2}
    \left[
    - (R_0\sin\phi+I_0\cos\phi)
        \sinh\left(\frac{\Delta\Gamma\,t}{2}\right)
    + (R_0\cos\phi-I_0\sin\phi)
        \sin\left(\Delta m\,t\right)
    \right]
  \nonumber \\
&& \qquad \qquad \qquad \qquad
  \Biggl.
  - \frac1{\left(R_{\|}^2+I_{\|}^2\right)\left(R_0^2+I_0^2\right)}
    \left(I_{\|}R_0-R_{\|}I_0\right)
    \left[ \cosh\left(\frac{\Delta\Gamma\,t}{2}\right)
         - \cos\left(\Delta m\,t\right)
    \right]
  \Biggr\}; \\
&& Im\left[\bar A_0^*(t) \bar A_{\|}(t)\right] =
\biggl[ 
   Re\left[A_0^*(0)A_{\|}(0)\right] \left( I_{\|}R_0-R_{\|}I_0 \right)
 + Im\left[A_0^*(0)A_{\|}(0)\right] \left( R_{\|}R_0+I_{\|}I_0 \right)
\biggr]
  \frac{e^{-\Gamma t}}{2} \times
  \nonumber \\
&& \qquad \qquad \qquad
  \Biggl\{
    \left[ \cosh\left(\frac{\Delta\Gamma\,t}{2}\right)
         + \cos\left(\Delta m\,t\right)
    \right]
  \Biggr. \nonumber \\
&& \qquad \qquad \qquad \qquad
  + \frac1{R_{\|}^2+I_{\|}^2}
    \left[
      (R_{\|}\cos\phi-I_{\|}\sin\phi)
        \sinh\left(\frac{\Delta\Gamma\,t}{2}\right)
    + (R_{\|}\sin\phi+I_{\|}\cos\phi)
        \sin\left(\Delta m\,t\right)
    \right]
  \nonumber \\
&& \qquad \qquad \qquad \qquad
  + \frac1{R_0^2+I_0^2}
    \left[
      (R_0\cos\phi-I_0\sin\phi)
        \sinh\left(\frac{\Delta\Gamma\,t}{2}\right)
    + (R_0\sin\phi+I_0\cos\phi)
        \sin\left(\Delta m\,t\right)
    \right]
  \nonumber \\
&& \qquad \qquad \qquad \qquad
  \Biggl.
  + \frac1{\left(R_{\|}^2+I_{\|}^2\right)\left(R_0^2+I_0^2\right)}
    \left(R_{\|}R_0+I_{\|}I_0\right)
    \left[ \cosh\left(\frac{\Delta\Gamma\,t}{2}\right)
         - \cos\left(\Delta m\,t\right)
    \right]
  \Biggr\} \nonumber \\
&& \qquad \qquad \qquad
- \biggl[ 
    Re\left[A_0^*(0)A_{\|}(0)\right]
        \left( R_{\|}R_0+I_{\|}I_0 \right)
  - Im\left[A_0^*(0)A_{\|}(0)\right]
        \left( I_{\|}R_0-R_{\|}I_0 \right)
  \biggr]
  \frac{e^{-\Gamma t}}{2} \times
  \nonumber \\
&& \qquad \qquad \qquad \qquad
  \Biggl\{
  \frac1{R_{\|}^2+I_{\|}^2}
    \left[
      (R_{\|}\sin\phi+I_{\|}\cos\phi)
        \sinh\left(\frac{\Delta\Gamma\,t}{2}\right)
    - (R_{\|}\cos\phi-I_{\|}\sin\phi)
        \sin\left(\Delta m\,t\right)
    \right]
  \Biggr. \nonumber \\
&& \qquad \qquad \qquad \qquad
  - \frac1{R_0^2+I_0^2}
    \left[
      (R_0\sin\phi+I_0\cos\phi)
        \sinh\left(\frac{\Delta\Gamma\,t}{2}\right)
    - (R_0\cos\phi-I_0\sin\phi)
        \sin\left(\Delta m\,t\right)
    \right]
  \nonumber \\
&& \qquad \qquad \qquad \qquad
  \Biggl.
  + \frac1{\left(R_{\|}^2+I_{\|}^2\right)\left(R_0^2+I_0^2\right)}
    \left(I_{\|}R_0-R_{\|}I_0\right)
    \left[ \cosh\left(\frac{\Delta\Gamma\,t}{2}\right)
         - \cos\left(\Delta m\,t\right)
    \right]
  \Biggr\}; \\
\label{conjugatere}
&& Re\left[\bar A_0^*(t) \bar A_{\bot}(t)\right] =
\biggl[ 
   Re\left[A_0^*(0)A_{\bot}(0)\right] \left( R_{\bot}R_0+I_{\bot}I_0 \right)
 - Im\left[A_0^*(0)A_{\bot}(0)\right] \left( I_{\bot}R_0-R_{\bot}I_0 \right)
\biggr]
  \frac{e^{-\Gamma t}}{2} \times
  \nonumber \\
&& \qquad \qquad \qquad
  \Biggl\{
    \left[ \cosh\left(\frac{\Delta\Gamma\,t}{2}\right)
         + \cos\left(\Delta m\,t\right)
    \right]
  \Biggr. \nonumber \\
&& \qquad \qquad \qquad \qquad
  - \frac1{R_{\bot}^2+I_{\bot}^2}
    \left[
      (R_{\bot}\cos\phi-I_{\bot}\sin\phi)
        \sinh\left(\frac{\Delta\Gamma\,t}{2}\right)
    + (R_{\bot}\sin\phi+I_{\bot}\cos\phi)
        \sin\left(\Delta m\,t\right)
    \right]
  \nonumber \\
&& \qquad \qquad \qquad \qquad
  + \frac1{R_0^2+I_0^2}
    \left[
      (R_0\cos\phi-I_0\sin\phi)
        \sinh\left(\frac{\Delta\Gamma\,t}{2}\right)
    + (R_0\sin\phi+I_0\cos\phi)
        \sin\left(\Delta m\,t\right)
    \right]
  \nonumber \\
&& \qquad \qquad \qquad \qquad
  \Biggl.
  - \frac1{\left(R_{\bot}^2+I_{\bot}^2\right)\left(R_0^2+I_0^2\right)}
    \left(R_{\bot}R_0+I_{\bot}I_0\right)
    \left[ \cosh\left(\frac{\Delta\Gamma\,t}{2}\right)
         - \cos\left(\Delta m\,t\right)
    \right]
  \Biggr\} \nonumber \\
&& \qquad \qquad \qquad
- \biggl[ 
    Re\left[A_0^*(0)A_{\bot}(0)\right] \left( I_{\bot}R_0-R_{\bot}I_0 \right)
  + Im\left[A_0^*(0)A_{\bot}(0)\right] \left( R_{\bot}R_0+I_{\bot}I_0 \right)
  \biggr]
  \frac{e^{-\Gamma t}}{2} \times
  \nonumber \\
&& \qquad \qquad \qquad \qquad
  \Biggl\{
  \frac1{R_{\bot}^2+I_{\bot}^2}
    \left[
      (R_{\bot}\sin\phi+I_{\bot}\cos\phi)
        \sinh\left(\frac{\Delta\Gamma\,t}{2}\right)
    - (R_{\bot}\cos\phi-I_{\bot}\sin\phi)
        \sin\left(\Delta m\,t\right)
    \right]
  \Biggr. \nonumber \\
&& \qquad \qquad \qquad \qquad
  + \frac1{R_0^2+I_0^2}
    \left[
      (R_0\sin\phi+I_0\cos\phi)
        \sinh\left(\frac{\Delta\Gamma\,t}{2}\right)
    - (R_0\cos\phi-I_0\sin\phi)
        \sin\left(\Delta m\,t\right)
    \right]
  \nonumber \\
&& \qquad \qquad \qquad \qquad
  \Biggl.
  + \frac1{\left(R_{\bot}^2+I_{\bot}^2\right)\left(R_0^2+I_0^2\right)}
    \left(I_{\bot}R_0-R_{\bot}I_0\right)
    \left[ \cosh\left(\frac{\Delta\Gamma\,t}{2}\right)
         - \cos\left(\Delta m\,t\right)
    \right]
  \Biggr\}; \\
\label{conjugateim}
&& Im\left[\bar A_0^*(t) \bar A_{\perp}(t)\right] =
\biggl[ 
   Re\left[A_0^*(0)A_{\perp}(0)\right] \left( I_{\perp}R_0-R_{\perp}I_0 \right)
 + Im\left[A_0^*(0)A_{\perp}(0)\right] \left( R_{\perp}R_0+I_{\perp}I_0 \right)
\biggr]
  \frac{e^{-\Gamma t}}{2} \times
  \nonumber \\
&& \qquad \qquad \qquad
  \Biggl\{
    \left[ \cosh\left(\frac{\Delta\Gamma\,t}{2}\right)
         + \cos\left(\Delta m\,t\right)
    \right]
  \Biggr. \nonumber \\
&& \qquad \qquad \qquad \qquad
  - \frac1{R_{\perp}^2+I_{\perp}^2}
    \left[
      (R_{\perp}\cos\phi-I_{\perp}\sin\phi)
        \sinh\left(\frac{\Delta\Gamma\,t}{2}\right)
    + (R_{\perp}\sin\phi+I_{\perp}\cos\phi)
        \sin\left(\Delta m\,t\right)
    \right]
  \nonumber \\
&& \qquad \qquad \qquad \qquad
  + \frac1{R_0^2+I_0^2}
    \left[
      (R_0\cos\phi-I_0\sin\phi)
        \sinh\left(\frac{\Delta\Gamma\,t}{2}\right)
    + (R_0\sin\phi+I_0\cos\phi)
        \sin\left(\Delta m\,t\right)
    \right]
  \nonumber \\
&& \qquad \qquad \qquad \qquad
  \Biggl.
  - \frac1{\left(R_{\perp}^2+I_{\perp}^2\right)\left(R_0^2+I_0^2\right)}
    \left(R_{\perp}R_0+I_{\perp}I_0\right)
    \left[ \cosh\left(\frac{\Delta\Gamma\,t}{2}\right)
         - \cos\left(\Delta m\,t\right)
    \right]
  \Biggr\} \nonumber \\
&& \qquad \qquad \qquad
- \biggl[ 
    Re\left[A_0^*(0)A_{\perp}(0)\right]
        \left( R_{\perp}R_0+I_{\perp}I_0 \right)
  - Im\left[A_0^*(0)A_{\perp}(0)\right]
        \left( I_{\perp}R_0-R_{\perp}I_0 \right)
  \biggr]
  \frac{e^{-\Gamma t}}{2} \times
  \nonumber \\
&& \qquad \qquad \qquad \qquad
  \Biggl\{
  \frac1{R_{\perp}^2+I_{\perp}^2}
    \left[
    - (R_{\perp}\sin\phi+I_{\perp}\cos\phi)
        \sinh\left(\frac{\Delta\Gamma\,t}{2}\right)
    + (R_{\perp}\cos\phi-I_{\perp}\sin\phi)
        \sin\left(\Delta m\,t\right)
    \right]
  \Biggr. \nonumber \\
&& \qquad \qquad \qquad \qquad
  + \frac1{R_0^2+I_0^2}
    \left[
    - (R_0\sin\phi+I_0\cos\phi)
        \sinh\left(\frac{\Delta\Gamma\,t}{2}\right)
    + (R_0\cos\phi-I_0\sin\phi)
        \sin\left(\Delta m\,t\right)
    \right]
  \nonumber \\
&& \qquad \qquad \qquad \qquad
  \Biggl.
  - \frac1{\left(R_{\perp}^2+I_{\perp}^2\right)\left(R_0^2+I_0^2\right)}
    \left(I_{\perp}R_0-R_{\perp}I_0\right)
    \left[ \cosh\left(\frac{\Delta\Gamma\,t}{2}\right)
         - \cos\left(\Delta m\,t\right)
    \right]
  \Biggr\}.
\end{eqnarray}
Similar formulas for $Re\left[\bar A_{\|}^*(0) \bar
A_{\bot}(0)\right]$ and $Im\left[\bar A_{\|}^*(0) \bar
A_{\bot}(0)\right]$ can be obtained from Eq.~(\ref{conjugatere}) and
(\ref{conjugateim}) by replacing ``$0$'' with ``$\|$'', respectively.

\end{appendix}

%%%%%%%%%%%%%%%%%%%%%%%%%%%%%%%%%%%%%%%%%%%%%%%%%%%%%%%%%%%%%%%

{\tighten

}


\begin{references}
\bibitem{DDLR96} A.S. Dighe, I. Dunietz, H.J. Lipkin, and
J.L. Rosner, Phys. Lett. B{\bf 369}, 144 (1996).

\bibitem{DDF99} A.S. Dighe, I. Dunietz, and R. Fleischer,
Eur. Phys. J. C{\bf 6}, 647 (1999).

\bibitem{KP92PRD} G. Kramer and W.F. Palmer, Phys. Rev. D{\bf 45}, 193
(1992).

\bibitem{KP92PLB} G. Kramer and W.F. Palmer, Phys. Lett. B{\bf 279},
 181 (1992).

\bibitem{KPS94} G. Kramer, W.F. Palmer, and H. Simma,
Nucl. Phys. {\bf B428}, 77 (1994).

\bibitem{KMP92} G. Kramer, T. Mannel, and W.F. Palmer, Z. Phys. C{\bf
55}, 497 (1992).

\bibitem{F99} R. Fleischer, Phys. Rev. D {\bf 60}, 073008 (1999).

\bibitem{LSS99} D. London, N. Sinha, and R. Sinha, Phys. Rev. D {\bf
60}, 074020 (1999).

\bibitem{CT99} C.W. Chiang and B. Tseng, hep-ph9905338.

\bibitem{CW99} C.W. Chiang and L. Wolfenstein, hep-ph/9911338 (to
appear in Phys. Rev. {\bf D}) (1999).

\bibitem{CY99} H.Y. Cheng and K.C. Yang, Phys. Rev. D {\bf 59},
092004 (1999).

\bibitem{PDG98} Particle Data Group, Eur. Phys. J. C{\bf 3} (1998).

\bibitem{CDF98} F. Abe et. al., CDF Collaboration,
Phys. Rev. Lett. {\bf 81}, 5513 (1998).

\bibitem{N92} Y. Nir, Lectures given at 20th Annual SLAC Summer
Institute on Particle Physics: The Third Family and the Physics of
Flavor, Stanford, CA, 13-24 Jul 1992.

\bibitem{W83} L. Wolfenstein, Phys. Rev. Lett. {\bf 51}, 1945 (1983).

\bibitem{FS78} D. Fakirov and B. Stech, Nucl. Phys. {\bf B133}, 315
(1978).

\bibitem{BSW85} M. Bauer, B. Stech and M. Wirbel, Z. Phys. {\bf C29},
637 (1985).

\bibitem{BSW87} M. Bauer, B. Stech and M. Wirbel, Z. Phys. {\bf C34},
103 (1987).

\bibitem{KG79} J. K\"{o}rner and G. Goldstein, Phys. Lett. {\bf 89B},
105 (1979).

\bibitem{B89} J.D. Bjorken, Nucl. Phys. Proc. Suppl. {\bf B11}, 325
(1989).

\bibitem{BHP96} T.E. Browder, K. Honscheid and D. Pedrini,
Annu. Rev. Nucl. Part. Sci. {\bf 46}, 395 (1996).

\bibitem{AYOPR93} R. Aleksan, A. Le Yaouanc, L. Oliver, O. P\`{e}ne
and Y.C. Raynal, Phys. Lett. {\bf B316}, 567 (1993).

\bibitem{BH92} J. Bijnens and F. Hoogeveen, Phys. Lett. {\bf B283},
434 (1992).

\bibitem{R90} J.L. Rosner, Phys. Rev. D {\bf 42}, 3732 (1990).

\bibitem{MRR91} T. Mannel, W. Roberts and Z. Ryzak, Phys. Rev. D {\bf
44}, 18 (1991).

\bibitem{BSS79} M. Bander, D. Silverman, and A. Soni,
Phys. Rev. Lett. {\bf 43}, 242 (1979).

\bibitem{NS97} M. Neubert and B. Stech, in {\it Heavy Flavours},
edited by A.J. Buras and M. Lindner, 2nd ed. (World Scientific,
Singapore, 1998), p. 294, hep-ph/9705292.

\bibitem{J90} W. Jaus, Phys. Rev. D. {\bf 41}, 3394 (1990).

\bibitem{CCW97} H.Y. Cheng, C.Y. Cheung, and C.W. Hwang, Phys. Rev. D
{\bf 55}, 1559 (1997).

\end{references}
\end{document}